\documentclass[amsmath,amssymb,aps,pra,floatfix,twocolumn]{revtex4-1}

\usepackage{color}
\usepackage{graphicx}
\usepackage{tikz}% drawing package
\usepackage{blindtext}
\usepackage{enumerate}
\usepackage{color}
\usepackage{slashed}

\begin{document}

\title{Interferometric modulation of  quantum cascade interactions}

\author{Stefano Cusumano}
\affiliation{NEST, Scuola Normale Superiore and Istituto Nanoscienze-CNR, I-56127 Pisa, Italy}
\email{stefano.cusumano@sns.it}

\author{Andrea Mari}
\affiliation{NEST, Scuola Normale Superiore and Istituto Nanoscienze-CNR, I-56127 Pisa, Italy}
%\email{andrea.mari@sns.it}

\author{Vittorio Giovannetti}
\affiliation{NEST, Scuola Normale Superiore and Istituto Nanoscienze-CNR, I-56127 Pisa, Italy}
%\email{vittorio.giovannetti@sns.it}

\begin{abstract}
We consider many-body quantum systems dissipatively coupled by a cascade network, {\it i.e.} a setup in which interactions are mediated by unidirectional environmental modes propagating through a linear optical interferometer. In particular we are interested in the possibility of inducing different effective  interactions by properly engineering an external dissipative network of beam-splitters and phase-shifters.
In this work we first derive the general structure of the master equation for a symmetric class of translation-invariant cascade networks. Then we show how, by tuning the parameters of the interferometer, one can exploit interference effects to tailor a large variety of many-body interactions.

\end{abstract}
\maketitle

\section{\label{sec:intro}Introduction}
Quantum cascade systems (QCSs) are particular physical configurations in which a quantum system can affect the dynamics of another system  but not {\it vice versa}.  This kind of asymmetric interaction is typical of experimental situations in which the coupling between the systems is not direct, but is instead mediated by external environmental modes which are forced to propagate along unidirectional channels: e.g. optical isolators or chiral bosonic channels~\cite{NATURE}.
QCSs have been theoretically and experimentally studied mainly within the field of quantum optics \cite{PhysRevLett.70.2273,PhysRevA.50.1792,PhysRevA.31.3761,PhysRevLett.70.2269,gardinerbook}, especially for dealing with the typical situation in which spatially separated quantum systems are connected by unidirectional laser beams. More recently,  the potential of QCSs has been theoretically investigated in very different contexts, such as: quantum state preparation and quantum computation~\cite{PhysRevLett.113.237203,PhysRevA.89.022335,PhysRevA.84.042341,1367-2630-14-6-063014}, chiral quantum networks~\cite{PhysRevA.91.042116,PhysRevLett.116.093601,PhysRevA.86.042306, Combes2017},  heat transmission~\cite{PhysRevA.91.022121, Mari2015}, {\it etc.}. Moreover also experimental implementations have been proposed, ranging from nanophotonic waweguides~\cite{Sollner:2015aa,1606.07466} to spin-orbit coupling~\cite{Petersen67}.

 \begin{figure}[!t]
\centering
\includegraphics[scale=0.50]{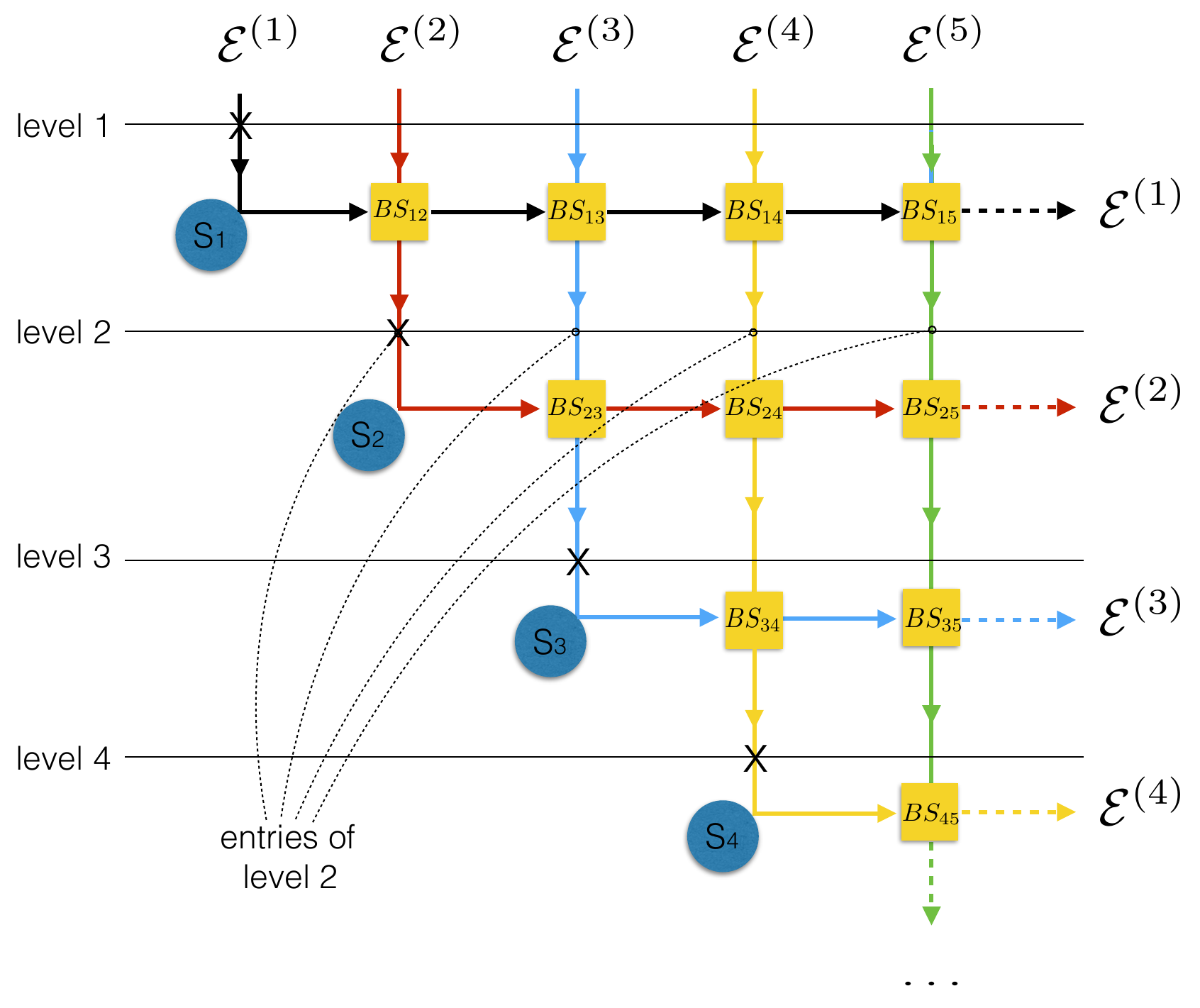}
\caption{Pictorial representation of the system studied: interactions between the quantum systems ${\cal S}:= \{ S_1,S_2, \cdots\}$ are mediated by a network of multi-mode bosonic chiral environmental channels $\mathcal{E} := \{ \mathcal{E}^{(1)},\mathcal{E}^{(2)},\cdots\}$ 
which interfere through a collection of  beam splitters $BS_{ij}$ (yellow squares in the figure) while progressing, from top to bottom, through the various levels of the network (thin horizontal lines). The X symbols in the figure identifies the first entry of the various levels.   }
\label{fig:system}
\end{figure}

From a theoretical point of view, the natural setting for studying QCSs is provided  by the theory of 
open quantum systems~\cite{breuer-petruccione,gardinerbook}.
A standard approach to study 
 QCSs is the so called LSH formalism~\cite{gough1,gough2,gough3,combes}.  It  is  based on an input-output description of  the couplings connecting the various components 
 of the system of interest ${\cal S}$. Accordingly it  
 is particularly suited to 
directly address the dissipative signals emerging from  ${\cal S}$, while it typically
requires a more elaborate analysis involving stochastic calculus to get the resulting master equation for  ${\cal S}$ alone.  
An alternative representation of QCS can be derived 
by adopting a  collisional model approach~\cite{prl108040401,jpbatmolopt45154003,cusumano_coll_model} which instead is directly focused on the
dynamics of ${\cal S}$. This is the route we follow in the present work. Specifically we focus on the  dynamics of a many-body quantum system ${\cal S}$ 
whose subparts are interconnected by signals that propagated unidirectionally through a complex  network composed by beams-splitter and phase shifter elements. 
Neglecting the delay times required by the  coupling signals from each 
controller element of ${\cal S}$ to its  controlled neighbours 
the dynamics  of  QCSs can be well approximated by 
  effective Born-Markov master equations whose generators exhibit a peculiar  structure that reflect the asymmetry of the associate interactions~\cite{1367-2630-14-6-063014}. 
  When casted in the standard Gorini, Kossakowski, Sudarshan, and Lindblad (GKSL) form~\cite{KOS,LIN,GO}, they produce effective coupling Hamiltonians  which, under proper conditions, possess special chiral symmetries~\cite{1367-2630-14-6-063014,cusumano_coll_model,arxiv160504312}. In a previous publication~\cite{cusumano_coll_model} we showed  how 
 interference effects in the propagation of the  signals could be used to to externally modulate the resulting QCS coupling among a limited number of sites, e.g. suppressing all but the first-neighbouring interactions.
 In the present paper we generalize this results by  showing that the same effect can be observed for an arbitrary number of sites if a proper network confirguration is adopted.  
More generally the configuration we analyze here, despite been relatively simple,  appears to be well suited to simulate a reach variety of dynamical behaviours  by externally acting on the system parameters.

The paper goes as follow:
in Sec.~\ref{sec:the_me} we show how to write the master equation for the network using the collisional model in~\cite{cusumano_coll_model} and highlight its main features. 
In particular Sec.~\ref{STANDARD} is devoted to   show how one can express the resulting equation in standard GKSL form~\cite{KOS,LIN,GO}; Sec.~\ref{sec:comp} instead provides 
the explicit computation of the system coupling constants; while Sec.~\ref{sec:loss} discusses how losses affecting the coupling signals can be included in the model. 
In Sec.~\ref{REGULAR} we focus on the case of regular networks which yields translational invariant  QCS couplings. 
In this context we also illustrate how the sites interactions can be tailored by exploiting  the interference effects associated with the propagation of the signals through the network.
In particular we identify the setting that allows one to eliminate all the couplings but those involving  first-neighboring sites. 
Finally in Sec.~\ref{sec:conclusions} we resume our results and draw our conclusions.

\begin{figure}[!t]
\centering
\includegraphics[scale=0.5]{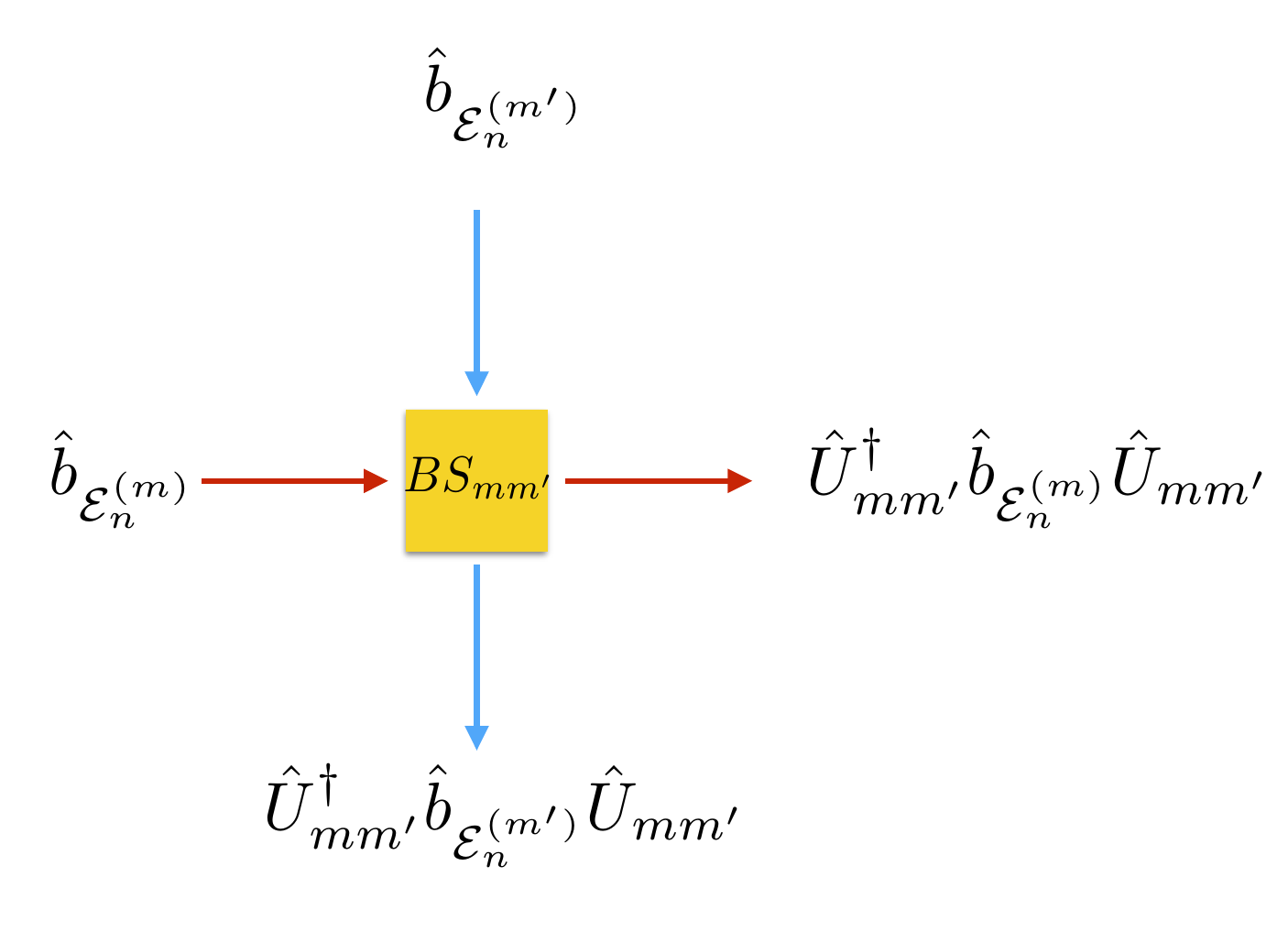}
\caption{Input-output mapping  in Heisenberg representation~(\ref{mapping})  induced by the beam splitter $BS_{mm'}$ that couples the channels 
$\mathcal{E}^{(m)}$ and $\mathcal{E}^{(m')}$ ($m<m'$).}
\label{fig:BS_action}
\end{figure}

\section{\label{sec:the_me}The master equation}
Consider the system depicted in Fig.~\ref{fig:system}. Here a set of  $M$ ordered quantum systems ${\cal S}:=\{ S_1, S_2, \cdots, S_M\}$ (e.g. optical cavities or two level systems) 
are connected via a network of mutually intercepting unidirectional channels $\mathcal{E}:=\{ \mathcal{E}^{(1)}, \mathcal{E}^{(2)}, \cdots, \mathcal{E}^{(M)}\}$ each represented by a collection of almost resonant, chiral bosonic environmental  modes  with annihilation operators $\{ \hat{b}_{ \mathcal{E}^{(m)}_n}\}_n$ fulfilling canonical commutation relations
  \begin{eqnarray}
\left[\hat{b}_{\mathcal{E}^{(m)}_n},\hat{b}_{\mathcal{E}^{(m')}_{n'}}^{\dagger}\right]=\delta_{mm'} \delta_{nn'}\;, \quad 
\left[\hat{b}_{\mathcal{E}_n^{(m)}},\hat{b}_{\mathcal{E}_{n'}^{(m')}}\right]=0\;,
\end{eqnarray}
the index $n$ referring to the mode degeneracy of each given channel. In our construction we assume these modes to enter the network as vacuum states and to propagate through an organized serie of
 beam-splitter transformations  which coherently mixes them. 
 Specifically for $m<m'$,   
   the channels $\mathcal{E}^{(m)}$ and $\mathcal{E}^{(m')}$ intercept at the beam splitter $BS_{mm'}$ described by the unitary transformation $\hat{U}_{m,m'}$ which, in Heisenberg representation,  induces the following tranformation: 
 \begin{widetext}    \begin{eqnarray} 
 \hat{b}_{\mathcal{E}^{(m)}_n} &\longrightarrow&   \hat{U}_{m,m'}^\dag \hat{b}_{\mathcal{E}^{(m)}_n} \hat{U}_{m,m'} = \sqrt{t_{m,m'} }\;   \hat{b}_{\mathcal{E}^{(m)}_n}  - i \sqrt{1-t_{m,m'}}  \; \hat{b}_{\mathcal{E}^{(m')}_n}  \;, \nonumber \\ 
 \hat{b}_{\mathcal{E}^{(m')}_n} &\longrightarrow&    \hat{U}_{m,m'}^\dag \hat{b}_{\mathcal{E}^{(m')}_n} \hat{U}_{m,m'} = e^{-i \phi_{m,m'}} \Big(\sqrt{t_{m,m'}} \;  \hat{b}_{\mathcal{E}^{(m')}_n}  - i \sqrt{1- t_{m,m'}} \;  \hat{b}_{\mathcal{E}^{(m)}_n} \Big)\;, \label{mapping} 
   \end{eqnarray}    
   \end{widetext} 
   with $t_{m,m'} \in [0,1]$ being the transmissivity of the device and ${\phi_{m,m'}} \in [0,2\pi]$ being the relative phase acquired by the two output modes (see also Fig.~\ref{fig:BS_action}).   Finally, as interaction  between ${\cal S}$ and ${\cal E}$, we take the  following exchange Hamiltonian \begin{eqnarray}
\hat{H}_{\mathcal{S},\mathcal{E}}=\sum_{m=1}^M \sum_n g_n\left(\hat{a}_{m}^\dagger \hat{b}_{\mathcal{E}^{(m)}_n}+\hat{a}_m\hat{b}_{\mathcal{E}^{(m)}_n}^\dagger\right)\;,
\end{eqnarray}
where $g_n$ are coupling constants and where for $m\in\{ 1, \cdots, M\}$, $\hat{a}_m$, $\hat{a}_m^\dagger$  are the  lowering and raising operators associated with subsystem $S_m$.
With the above premises, the temporal evolution of the reduced density matrix $\hat{\rho}$ of 
the systems ${\cal S}$ can be derived by enforcing proper Born-Markov approximations.  The resulting master equation takes the form 
\begin{eqnarray}
\label{eq:the_master_equation}
\frac{\partial\hat{\rho}}{\partial t}=\sum_{m}\mathcal{L}_m(\hat{\rho})+\sum_{m'> m} \mathcal{D}_{m\rightarrow m'}(\hat{\rho})\;,
\end{eqnarray}
with  the super-operators $\mathcal{L}_m$ and $\mathcal{D}_{m\rightarrow m'}$ describing respectively  local dissipation terms and cascade interactions mediated by the chiral modes, i.e.
\begin{eqnarray}
\label{eq:loc_term_zero_t}
\mathcal{L}_m(\cdots)=\frac{\gamma}{2}\left(2\hat{a}_m(\cdots)\hat{a}_m^\dag-\left[\hat{a}_m^\dag\hat{a}_m,\cdots\right]_{+}\right), 
\end{eqnarray}
and, for $m'>m$, 
\begin{eqnarray} 
&&\mathcal{D}_{m\rightarrow m'}(\cdots)=\label{eq:int_term_path}\\
\nonumber
&&\qquad \gamma\left(\zeta_{m,m'}\hat{a}_{m}\Big[\cdots,\hat{a}_{m'}^\dag\Big]_{-}+\zeta_{m,m'}^{*}\Big[\hat{a}_{m'},\cdots\Big]_{-}\hat{a}_{m}^\dag\right)\;,
\end{eqnarray}
where   $[\cdots,\cdots]_-$ and  $[\cdots,\cdots]_+$  represent the commutator and the anti-commutator, respectively. In these expressions
the parameter $\gamma$ sets the time-scale of the dissipation process. In the standard derivation of ME it originates from a particular combination of the coupling strength and the bath spectral density \cite{breuer-petruccione}, while in the formalism of Ref.~\cite{cusumano_coll_model} it can be expressed as the limit 
 $\gamma=\lim_{\Delta t\rightarrow0}g_n^2\Delta t$ with
$\Delta t$  being the collision time that rules the interaction between ${\cal S}$ and the modes of the channel.  As we shall discuss in the next section,
the complex coefficients $\zeta_{m,m'}$ depend instead upon the transmissivities and phases of beam-splitter that form the network.
This means that  modifying these parameters it is possible to tune the strength of the QCS interactions 
simulating  a rich variety of effective dynamics: e.g. realizing a chain of cascaded systems where only first-neighbor interactions are present, or only second-neighbor interactions and so on.
Furthermore, acting on the phase shifts one could also think of combinations of these situations, considering for instance a system where both first- and second-neighbor interactions are present with tunable  relative strength.

\subsection{GKSL standard form} \label{STANDARD} 

As discussed in Ref~\cite{cusumano_coll_model} the master equation (\ref{eq:the_master_equation}) can be equivalently casted into a standard GKSL form~\cite{KOS,LIN,GO}  which,
beside a purely dissipative contributions mediated by a collection of multipartite Lindblad operators $\hat{L}_i$ with corresponding rates $\gamma_i$, exhibits  an effective coupling Hamiltonian $\hat{H}_{eff}$, i.e. 
\begin{eqnarray}
\frac{\partial\hat{\rho}}{\partial t}=-i[\hat{H}_{eff} ,\hat{\rho}]_{-}+\sum_{i}  \frac{\gamma_i}{2}\left(2\hat{L}_i\hat{\rho}\hat{L}_i^\dag-\left[\hat{L}_i^\dag\hat{L}_i,\hat{\rho}\right]_{+}\right).
\nonumber \\ \label{eq:lindblad}
\end{eqnarray}
For the case under examination the resulting $\hat{H}_{eff}$ admits a simple expression as a sum of the following  two-body coupling terms 
\begin{eqnarray}
\hat{H}_{eff} =-\frac{i}{2} \sum_{m'>m} \left(\zeta_{m,m'}\; \hat{a}_{m}\hat{a}_{m'}^\dag-h.c.\right),
\label{eq:chiral_hamiltonian}
\end{eqnarray}
which, similarly to the  super-operators ${\cal D}_{m\rightarrow m'}$, exhibit strengths which are  mediated by the coupling constants
$\zeta_{m,m'}$. 
As discussed in Appendix~\ref{STAND}  the rates $\gamma_i$ are instead provided by the eigenvalues of the $M\times M$ Hermitian matrix $\Theta$ of elements 
\begin{eqnarray} \label{THETA1} 
\Theta_{m,m} &=& \gamma \;, \\ 
\Theta_{m,m'} &=&\Theta_{m',m}^* =  \gamma \; \zeta_{m,m'}  \qquad  \forall m'>m\;, \label{THETA2} 
\end{eqnarray} 
while the corresponding operators $\hat{L}_i$ are finally obtained as the following linear combinations of the annihilation operators $\hat{a}_m$
\begin{eqnarray} \label{DEFLf} 
\hat{L}_i = \sum_{m} w_{m,i}^* \hat{a}_{m}\;, \end{eqnarray} 
the  amplitudes $w_{m,i}$ being  the elements of the $M \times M$ unitary matrix which diagonalizes $\Theta$ via the identity 
$\gamma_i = \sum_{k,k'} w^*_{m,i} \Theta_{m,m'} w_{m',i}$.
As evident from the above expressions a closed expression for  $\hat{L}_i$ and $\gamma_i$ cannot be 
explicitly given for  an arbitrary choice of the network setting (see however
the next section for some special cases which admit a simple representation). However Eq.~(\ref{DEFLf}) makes it clear that in general the $\hat{L}_i$ will be not local, 
inducing cooperative emission processes that are somehow reminding us of Dicke-superradiance~\cite{DICKE}. 

%%%%%%
 \begin{figure}[!t]
\centering
\includegraphics[scale=0.4]{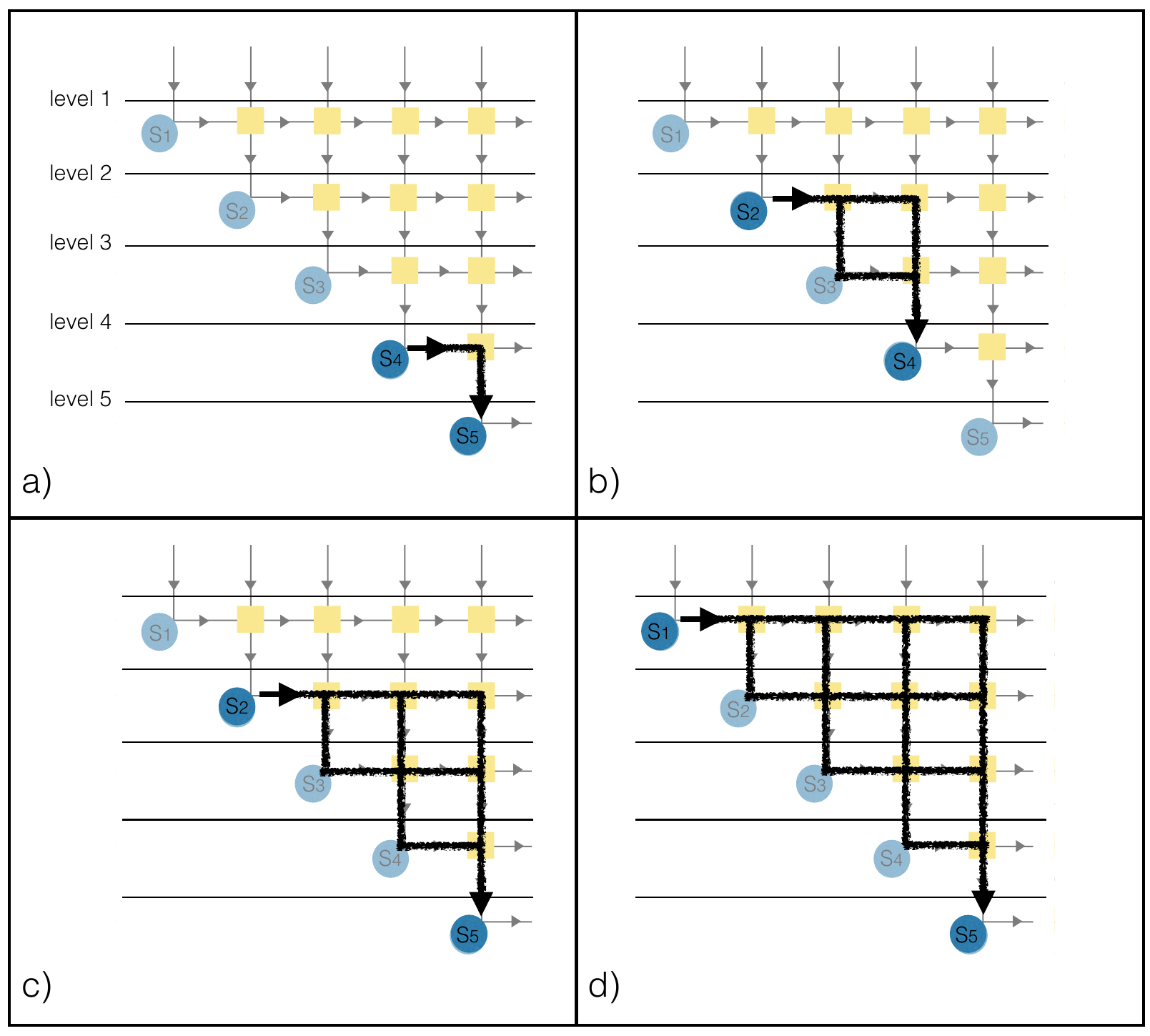}
\caption{Schematic representation of the paths (bold curves) that contribute to the coupling constants $\zeta_{m,m'}$ -- see Eq.~(\ref{DEF3}).
In particular:  panel a) shows the single path which enter into the definition of the coupling $\zeta_{4,5}$ between $S_4$ and $S_5$; panel b) those of 
$S_2$ and $S_4$; c) those of $S_2$ and $S_5$; d) those of $S_1$ and $S_5$.   } 
\label{figurenew5}
\end{figure}
%%%%%%%%

\subsection{Computing the couplings constants}\label{sec:comp}

In the previous section we have seen that  the couplings between the subsystems of the model are mediated by the  constants $\zeta_{m,m'}$ appearing in Eqs.~(\ref{eq:int_term_path}) and (\ref{eq:chiral_hamiltonian}). 
Following the derivation of  Ref~\cite{cusumano_coll_model} these can be computed as 
\begin{eqnarray} 
\zeta_{m,m'}=  \mbox{Tr}\Big[  \hat{b}_{\mathcal{E}^{(m')}_n}  {\cal M}^{(m'-1\leftarrow m)}_{\cal E} \Big( \hat{b}^\dag_{\mathcal{E}^{(m)}_n}  {\cal M}^{(m-1\leftarrow 1)} _{\cal E} (\eta_{\cal E})\Big) \Big]\;, 
\nonumber \\  \label{DEF1} 
\end{eqnarray} 
where $\eta_{\cal E}$ is the initial state of chiral channels,  while  ${\cal M}^{(m_2\leftarrow m_1)} _{\cal E}$ is the physical transformation that, in the absence of the interactions with the subsystems ${\cal S}$, describes the evolution of such state from the level $m_1$ of the network to the level $m_2 >  m_1$ (see Fig.~\ref{fig:system}). It can be expressed in terms of the ordered product of the transformations $BS_{m,m'}$ located between such levels. Specifically indicating with 
\begin{eqnarray} \hat{V}_m &=&  \cdots \hat{U}_{m,m+3}  \hat{U}_{m,m+2}  \hat{U}_{m,m+1} \;, \label{qui} 
\end{eqnarray} 
 the product of the  beam splitter unitary operators~(\ref{mapping}) that couples the channel ${\cal E}^{(m)}$  with the subsequent ones,
 and with 
 \begin{eqnarray} \hat{V}_{m_2 \leftarrow m_1}  &=& \hat{V}_{m_2} \cdots \hat{V}_{m_1+1}  \hat{V}_{m_1}  \;, \label{qui2} 
\end{eqnarray} 
the ordered product of such terms from $m_1$ to $m_2>m_1$, 
 we can express ${\cal M}^{(m_2\leftarrow m_1)} _{\cal E}$ as 
 \begin{eqnarray} 
{\cal M}^{(m_2\leftarrow m_1)} _{\cal E}: = \hat{V}_{m_2 \leftarrow m_1}  (\cdots)   \hat{V}_{m_2 \leftarrow m_1}^{\dag} \;.
\end{eqnarray}
Exploiting hence the composition rule
\begin{eqnarray} 
\hat{V}_{m_3 \leftarrow m_2}  \hat{V}_{m_2 \leftarrow m_1}  =  \hat{V}_{m_3 \leftarrow m_1} \;,
\end{eqnarray} 
which holds for all $m_3>m_2>m_1$ we can then rewrite Eq.~(\ref{DEF1})  as
\begin{eqnarray} 
\zeta_{m,m'}=  \mbox{Tr}[  \hat{c}_{\mathcal{E}^{(m')}_n}  \hat{c}^\dag_{\mathcal{E}^{(m)}_n}  \eta_{\cal E} ]\;, 
 \label{DEF11} 
\end{eqnarray} 
with $\hat{c}_{\mathcal{E}^{(m)}_n}$  being the annihilation operator $\hat{b}_{\mathcal{E}^{(m)}_n}$ evolved in Heisenberg picture from level $1$ to level $m$ of the network, i.e. 
\begin{eqnarray} 
\hat{c}_{\mathcal{E}^{(m)}_n}  := \hat{V}_{m-1 \leftarrow 1}^\dag \; \hat{b}_{\mathcal{E}^{(m)}_n} \;  \hat{V}_{m-1 \leftarrow 1} \;,
\end{eqnarray} 
the case $m=1$ being included by  identifying $ \hat{V}_{0 \leftarrow 1}$ with the identity operator so that $\hat{c}_{\mathcal{E}^{(1)}_n}  = \hat{b}_{\mathcal{E}^{(1)}_n}$\cite{NOTA}. 
For instance for $m=2$ we have 
\begin{eqnarray} 
\hat{c}_{\mathcal{E}^{(2)}_n}  &=&  \hat{U}^\dag_{12}  \hat{b}_{\mathcal{E}^{(2)}_n}  \hat{U}_{12} \\ &=&\nonumber 
e^{-i \phi_{12}} \Big(\sqrt{t_{12}} \;  \hat{b}_{\mathcal{E}^{(2)}_n}  - i \sqrt{1- t_{12}} \;  \hat{b}_{\mathcal{E}^{(1)}_n} \Big) \;, 
\end{eqnarray} 
where  we use Eq.~(\ref{mapping}) and the fact that, for $m>3$, $ \hat{b}_{\mathcal{E}^{(2)}_n}$ commutes with the operators $\hat{U}_{1m}$.
Similarly for $m=3$ we get 
\begin{widetext} 
\begin{eqnarray} 
\hat{c}_{\mathcal{E}^{(3)}_n}  &=&\hat{U}^\dag_{12}   \hat{U}^\dag_{13} \hat{U}^\dag_{23}   \hat{b}_{\mathcal{E}^{(3)}_n}    \hat{U}_{23}  \hat{U}_{13}  \hat{U}_{12} \nonumber \\ &=&  e^{-i  \phi_{23}}  \Big[ e^{-i \phi_{13} } \sqrt{t_{13} t_{23}} \; \hat{b}_{\mathcal{E}^{(3)}_n} 
 +  \Big(-i e^{-i \phi_{12} } \sqrt{t_{12}(1-t_{23})} -e^{-i \phi_{13}} \sqrt{(1-t_{12})(1-t_{13})t_{23}}\Big) \; \hat{b}_{\mathcal{E}^{(2)}_n} \nonumber \\&&\qquad
\qquad 
+ \Big( -e^{-i \phi_{12}} \sqrt{(1-t_{12})(1-t_{23})} - i e^{-i \phi_{13} } \sqrt{t_{12}(1-t_{13})t_{23}}\Big) \; \hat{b}_{\mathcal{E}^{(1)}_n}\Big] \;.
\end{eqnarray} 
\end{widetext} 
By closed inspection of the above expressions one can verify that 
 for generic $m$ the operator $\hat{c}_{\mathcal{E}^{(m)}_n}$ can be written as 
\begin{eqnarray}  \label{DEFC} 
\hat{c}_{\mathcal{E}^{(m)}_n}  &:=& \sum_{k=1}^m A^{(k)}_{m\leftarrow 1 } \; \hat{b}_{\mathcal{E}^{(k)}_n} \;, 
\end{eqnarray} 
where for $k\leq m$, the complex coefficients $A^{(k)}_{m\leftarrow 1 }$ are the probability amplitudes obtained 
by coherently summing over all the paths which brings the input mode $\hat{b}_{\mathcal{E}^{(k)}_n}$ from level $1$ to the first entry (see Fig.~\ref{fig:system}) of level $m$ of the network. Accordingly we get  
\begin{eqnarray}
\zeta_{m,m'}&=&\sum_{k'=1}^{m'} \sum_{k=1}^m  A_{m'\leftarrow 1 }^{(k')}\;  \left[A_{m\leftarrow 1 }^{(k)}\right]^*
\; \mbox{Tr}[ \hat{b}_{\mathcal{E}^{(k')}_n}  \hat{b}^\dag_{\mathcal{E}^{(k)}_n}  \eta_{\cal E}]  \;. \nonumber \\ 
\end{eqnarray} 
A further simplification occurs in the case where the input modes of the channels ${\cal E}$ are initialized into a collection of zero-mean factorized states. In particular
assuming $\eta_{\cal E}$ to be the vacuum state $|\O\rangle$ we arrive to 
\begin{eqnarray}
\zeta_{m,m'} \label{DEF3} 
&=&\sum_{k=1}^{m}    A^{(k)}_{m'\leftarrow 1 }\;  \left[A^{(k)}_{m\leftarrow 1 }\right]^*\;,  \qquad (m'>m)\;. 
\end{eqnarray} 
where we used the fact that \begin{eqnarray} \langle \O | \hat{b}_{\mathcal{E}^{(k')}_n}  \hat{b}^\dag_{\mathcal{E}^{(k)}_n}  |\O\rangle =\delta_{kk'}\;, \label{FACT} \end{eqnarray}
with $\delta_{kk'}$ being the Kronecker delta. 
As a matter of fact this
 is not the end of the story. Indeed exploiting the properties of the amplitudes $A_{m\leftarrow 1}$, Eq.~(\ref{DEF3}) can be equivalently casted into the following
extremely compact form 
\begin{eqnarray}
\zeta_{m,m'} \label{DEF3} 
&=&  A_{m'\leftarrow m }^{(m)}\;,  \qquad (m'>m)\;,
\end{eqnarray} 
showing that for the case of vacuum input modes, the coupling coefficients coincides with the probability amplitudes associated with the
propagation of signals from the node $S_m$ of the network to the node $S_{m'}$ -- see Fig.~\ref{figurenew5}. 
The easiest way to derive Eq.~(\ref{DEF3}) consists in going back to Eq.~(\ref{DEF11}) and using the fact that  vacuum states are invariant under the action of the beam splitter
transformations $\hat{U}_{m,m'}$ and, of course, of their concatenations (\ref{qui}) and (\ref{qui2}). 
Accordingly we can write 
\begin{eqnarray} 
\zeta_{m,m'}&=& \langle \O| \Big( \hat{V}_{m'-1\leftarrow m}^\dag  \hat{b}_{\mathcal{E}^{(m')}_n} \hat{V}_{m'-1\leftarrow m}  \Big) \hat{b}^\dag_{\mathcal{E}^{(m)}_n}  |\O\rangle \\ \nonumber 
&=& \sum_{k=m}^{m'} A^{(k)}_{m'\leftarrow k } \;  \langle \O|  \hat{b}_{\mathcal{E}^{(k)}_n}  \hat{b}^\dag_{\mathcal{E}^{(m)}_n}  |\O\rangle= A_{m'\leftarrow m }^{(m)}\;,
\end{eqnarray} 
where we used Eq.~(\ref{FACT}) and the identity 
\begin{eqnarray} 
 \hat{V}_{m'-1\leftarrow m}^\dag \;  \hat{b}_{\mathcal{E}^{(m')}_n} \; \hat{V}_{m'-1\leftarrow m}  =  \sum_{k=m}^{m'} A^{(k)}_{m'\leftarrow k } \; \hat{b}_{\mathcal{E}^{(k)}_n}\;, 
 \end{eqnarray} 
 which generalizes Eq.~(\ref{DEFC}).

 %%%%%%
 \begin{figure}[!t]
\centering
\includegraphics[scale=0.4]{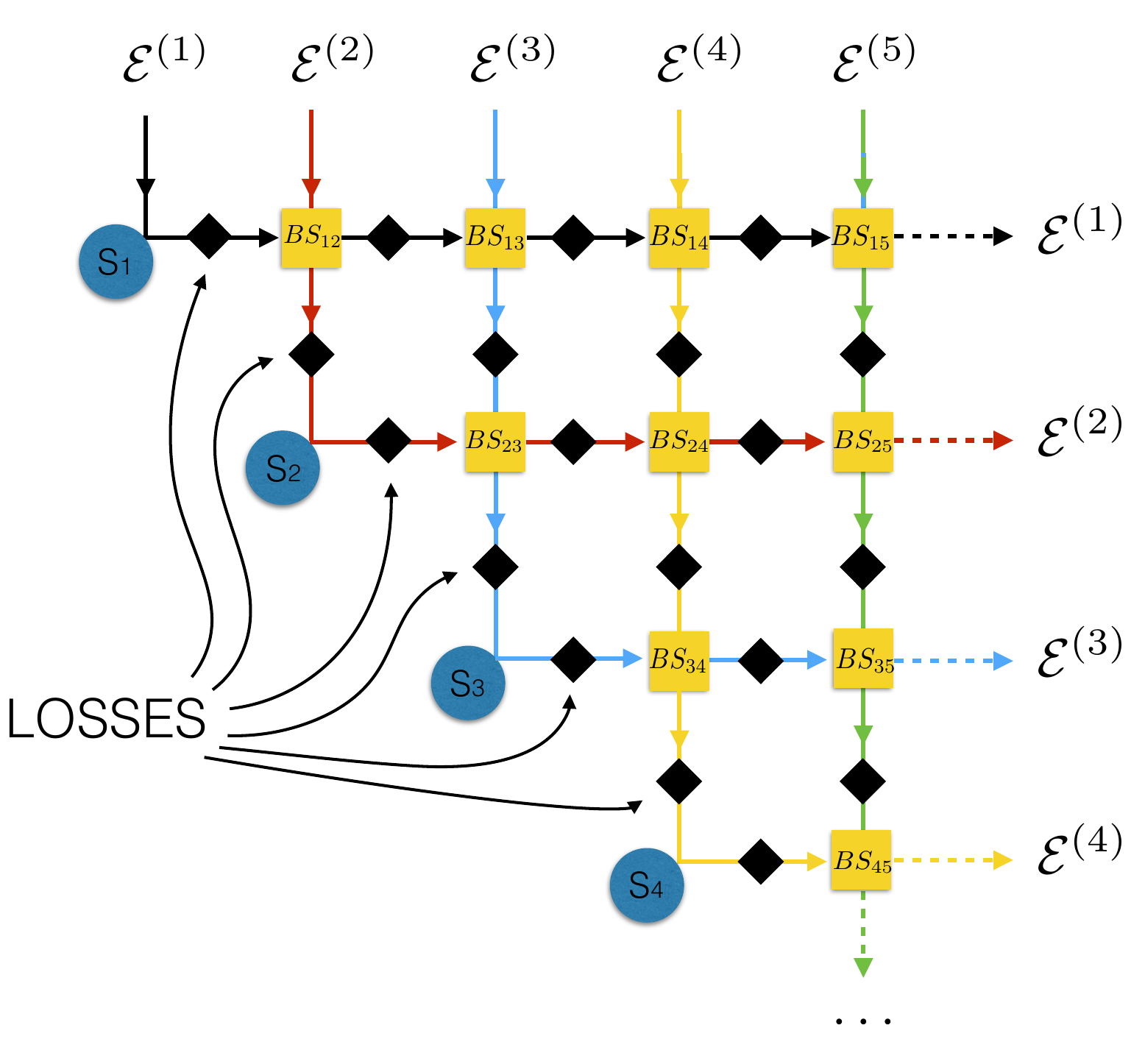}
\caption{Schematic representation of the  model in the presence of loss: the signal from the subsystems along the network have probability $\nu$ of being lost, i.e.
a probability $1-\nu$ of continuing their journey in the interferometric network.    } 
\label{figloss}
\end{figure}
%%%%%%%%

 \subsection{Losses in the network}  \label{sec:loss}
 In the previous analysis we have implicitly assumed that during their propagation along the network the coupling signals do not experience losses.
 The formalism however can also accomodate for these detrimental effects~\cite{prl108040401} by properly including them, as well as other form of noise that may tamper the model, into the definition of the maps ${\cal M}^{(m_2\leftarrow m_1)}_{\cal E}$
 of Eq.~(\ref{DEF1}). For instance, let us assume that each of the path that compose  the network is characterized by a probability $\nu \in [0,1]$ of losing the 
 signals which travel them -- see Fig.~\ref{figloss}. Then the new coupling constants $\zeta_{m,m'}$ entering Eqs.~(\ref{eq:the_master_equation})-(\ref{eq:chiral_hamiltonian}) acquire an 
extra  factor  which is exponentially decreasing with the sites distance $m'-m$, i.e.
 \begin{eqnarray} \label{LOSSY} 
\zeta_{m,m'}^{(\rm loss)} &:=& (1-\nu)^{m'-m} A_{m'\leftarrow m }^{(m)}\;,
\end{eqnarray} 
with $A_{m'\leftarrow m }^{(m)}$ being the probabilities amplitudes of the lossless regime ($\nu=0$). The easiest way to verify Eq.~(\ref{LOSSY}) is by modelling the losses via the
action of extra beam splitters of transmissivities $\sqrt{1-\nu}$, placed along the network in correspondence of the black elements of Fig.~\ref{figloss}, and coupling the signals with extra environmental degree of freedom initialized into the  vacuum.

\section{Regular network case} \label{REGULAR}
 \begin{figure}[!t]
\centering
\includegraphics[scale=0.7]{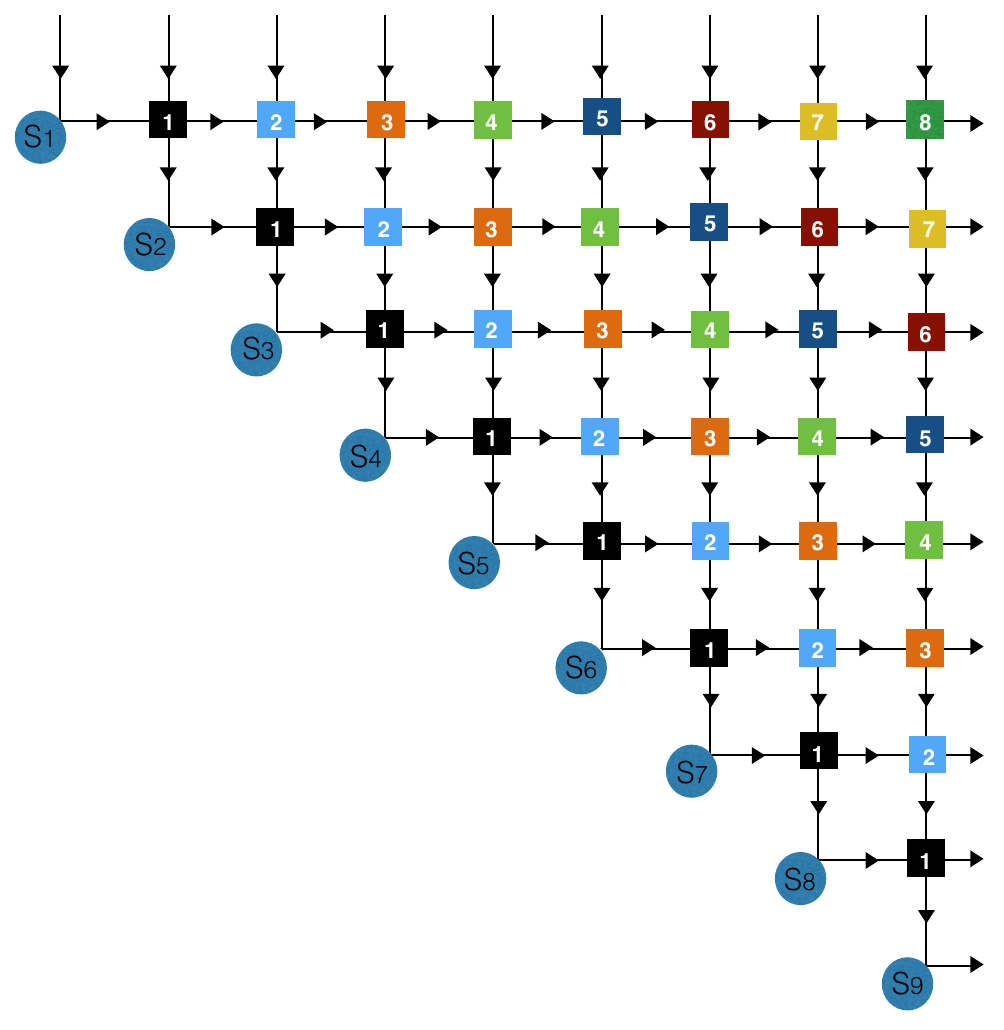}
\caption{Regular network case: as indicated by the labels the beam splitters (square elements of the figure) 
laying on the same diagonal of the network are assumed to be identical. 
For instance 
the elements $BS_{12}$,  $BS_{23}$, 
$BS_{34}$, $BS_{45}$, ... possess the same transmissivity $\tau_{k=1}$ and the same relative phase $\phi_{k=1}$, see Eq.~(\ref{condition}); similarly 
the elements $BS_{13}$,  $BS_{24}$, 
$BS_{35}$, $BS_{46}$, ... are characterized  instead  by transmissivity $\tau_{k=2}$ and by relative phase $\phi_{k=2}$.} 
\label{figurenew3}
\end{figure}
 In this section we focus on the special case of the regular network depicted in Fig.~\ref{figurenew3} where
 the beam splitters of the model  are organized in groups of identical elements, each group laying on the same diagonal of the network. 
Accordingly the transformations induced by $BS_{12}$, $BS_{23}$, $BS_{34}$, $\cdots$, $BS_{m,m+1}$, $\cdots$  which mediate the interactions among first neighboring channels
 are now assumed to be identical. Similarly the transformations associated with the beam splitters $BS_{13}$, $BS_{24}$, $BS_{35}$, $\cdots$, $BS_{m,m+2}$, $\cdots$  which instead mediate the interactions among second neighboring channels are also assumed to be identical. 
At mathematical level the above structure can be enforced by simply imposing 
 the symmetry 
 $\hat{U}_{m,m'} =  \hat{U}_{m+1,m'+1}$ for all   $m$,$m'$
 to the transformations~(\ref{mapping}), or equivalently by forcing the transmissivities and the relative phases of the model to obey the following constraint
 \begin{eqnarray} 
 t_{m,m+k} = \tau_{k} \;,  \qquad 
 \phi_{m,m+k} = \phi_{k} \;, \quad \forall m, k \label{condition} 
 \end{eqnarray} 
 where for $k=1,2, \cdots$,  $\tau_k$ and $\phi_k$ are assigned parameters.  
 Under these special conditions 
the coupling constants~(\ref{DEF3}) become explicitly invariant under translation of the indexes, i.e. 
\begin{eqnarray}
\zeta_{m,m'} \label{DEFtrans3} 
&=& \zeta_{1,m'-m+1}  \;,
\end{eqnarray} 
for all $m' >m$. This allows us to rewrite  the ME as 
\begin{eqnarray}
\label{eq:the_master_equation11}
\frac{\partial\hat{\rho}}{\partial t}=\sum_{m}\mathcal{L}_m(\hat{\rho})+ \sum_{k\geq 1}  {\cal D}_k \;, 
\end{eqnarray}
with the super-operators $\mathcal{D}_{k}$ being translationally invariant 
\begin{eqnarray} 
&&\mathcal{D}_{k}(\cdots)=\label{eq:int_term_path11212}\\
\nonumber
&&\qquad \gamma \sum_{m} \left(\xi_{k}\hat{a}_{m}\Big[\cdots,\hat{a}_{m+k}^\dag\Big]_{-}+\xi_{k}^{*}\Big[\hat{a}_{m+k},\cdots\Big]_{-}\hat{a}_{m}^\dag\right)\;,
\end{eqnarray}  
with 
coupling strengths 
 \begin{eqnarray}
 \label{eq:xi_k_def}
  \xi_{k} :=  \zeta_{1,k+1} = \;  A_{k+1\leftarrow 1 }^{(1)}\;,
\end{eqnarray} 
which in the presence of losses become
 \begin{eqnarray}
 \label{eq:xi_k_defLOSS}
  \xi_{k}^{(\rm loss)} =(1-\nu)^{k} \;  A_{k+1\leftarrow 1 }^{(1)}\;,
\end{eqnarray} 
see Eq.~(\ref{LOSSY}).
At the level of the effective
Hamiltonian this corresponds to have \begin{eqnarray}
\label{EFFECTIVE} 
\hat{H}_{eff} =  \sum_{k\geq 1} \hat{H}_{k}\;,  
\end{eqnarray}
where again for $k\geq 1$, $\hat{H}_{k}$ is a translationally invariant term involving sites which are $k$-th neighbouring  
 \begin{eqnarray} 
 \label{eq:H^{(k)}}
 \hat{H}_{k} := -\frac{i\gamma}{2}  \sum_{m} \left( \xi_k \;   \hat{a}_{m}\hat{a}_{m+k}^\dag-h.c.\right) \;.
 \end{eqnarray}

By construction, for a given value of $k$, the quantity $\xi_k$ depends upon the parameters $\tau_{k'}$, and $\phi_{k'}$ with $k'\leq k$. 
Accordingly by properly tuning such terms we can change the many-body structure of the coupling. 
Unfortunately the explicit functional dependence of $\xi_{k}$ upon the system parameter is in general  rather complex. 
Yet in the next subsections we shall analyze some special cases which admit explicit analytic solution.

\subsection{\label{sec:examples}Finite size networks }

\begin{figure}[!t]
\centering
\includegraphics[scale=.5]{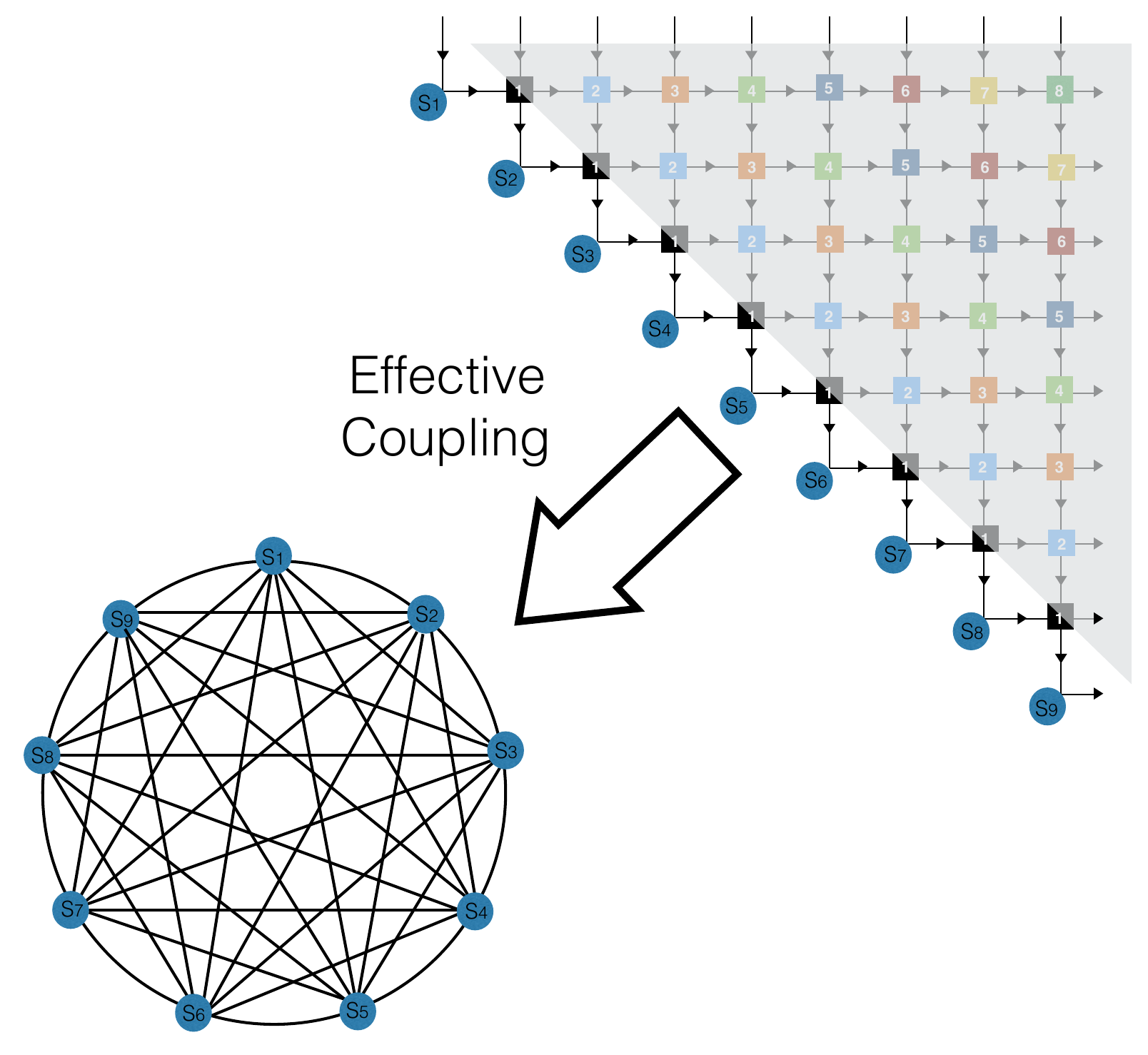}
\caption{Setting the transmissivity $\tau_{1}=0$ the sites are coupled via a single chiral channel. This induces an effective coupling between the sites mediated by a fully connected Hamiltonian with coupling strengths whose intensity is independent from the distance between the various elements.}
\label{figurenew4}
\end{figure}

Assume that the transmissivities of the beam splitters $BS_{m,m+1}$  which  couple  first neighboring channels is zero ($\tau_1=0$), see Fig.~\ref{figurenew4}.
In this case the network splits into two independent parts: the first, composed by the  chiral channel ${\cal E}^{(1)}$ linking all the sites of the model, the second
composed by the remaining environmental modes which instead do not interact with ${\cal S}$.  Under this assumption 
the coupling strengths $\xi_k$ can be easily computed. In the absence of losses during the propagation ($\nu=0$), up to an irrelevant phase, 
they have the same
intensity independently from the value of $k$, i.e. 
 \begin{eqnarray} 
  \xi_{k} =  (-i  e^{-i \phi_1})^k=  e^{-i k (\phi_1+\pi/2)} \;.\label{asdfas} 
\end{eqnarray} 
As a consequence 
the  effective Hamiltonian~(\ref{EFFECTIVE}) is fully connected with uniform coupling strengths, meaning that any given site interacts with all the others independently from their
relative distance, i.e.
\begin{eqnarray} 
\hat{H}_k&=&-\frac{i\gamma}{2}\sum_m  \left({  e^{-i k (\phi_1+\pi/2)} \hat{a}}_m {\hat{a}}_{m+k}^\dag-h.c. \right) \nonumber \\
&=&-\frac{i\gamma}{2}\sum_m\left({\hat{d}}_m {\hat{d}}_{m+k}^\dag-h.c. \right)\;, \label{QUESTAQUI} 
\end{eqnarray}
where the second identity explicitly shows that the phase $\phi_1+\pi/2$ is irrelevant as, for all $k$, it  can  be reabsorbed into the system operators, i.e.   $\hat{d}_m={\hat{a}}_me^{i m(\phi_1+\pi/2)}$ -- the remaining contributions of the ME being unaffected by the transformation.
An analogous behavior is also observed when we take  $\tau_1$ generic but assume that  all the other beam splitters of the network have unitary transmissivities, i.e. 
$\tau_k =1$ for all $k\geq 2$. Under this assumption the signals which get transmitted through the first line of beam splitter never have a chance of interacting with 
the subsystems. Accordingly the coupling constants become exponentially depressed,  \begin{eqnarray} 
  \xi_{k} = e^{-i k (\phi_1+\pi/2)} (1-\tau_1)^{k/2} \;, \label{asdfas111} 
\end{eqnarray} 
similarly to what one would observe for the case $\tau_1=0$ in the presence of
losses~(\ref{eq:xi_k_defLOSS}) 
 (the correspondence being exact by identifying  $\sqrt{1-\tau_1}$ with $1-\nu$). 
%%%%%%
%%%%

\begin{figure}[!t]
\centering
\includegraphics[scale=0.27]{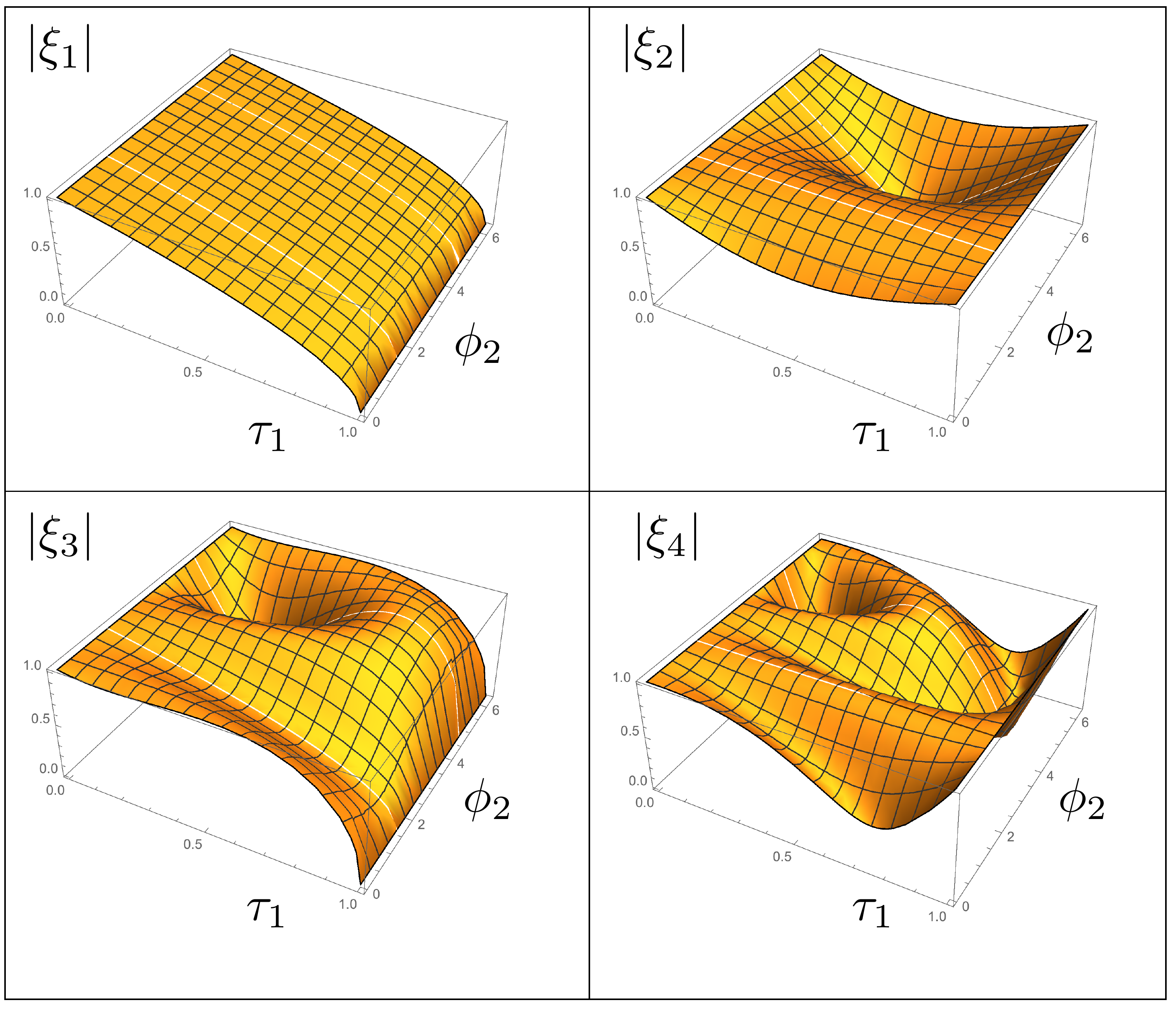}
\caption{Plot of the modulus of the first four coupling constants $\xi_k$ for the case of a regular network with $\tau_2=0$ as a function of $\tau_1$ and $\phi_2$. Notice that for $\tau_1=0$
all the coupling constants have modulus 1  in agreement with (\ref{asdfas}). For $\tau_1=1$ instead the odd terms nullify, while the even one maintain maximum value, see Eq.~(\ref{eq:odd_even_interactions}).
In all the plots we set $\phi_1=0$. 
} 
\label{figurenuovissima}
\end{figure}

A generalization of Eq.~(\ref{asdfas}) 
 is obtained considering the case where 
for  a given  $K\geq 2$ integer, the transmissivities of the beam splitters $BS_{m,m+K}$ are zero, i.e.  $\tau_K=0$. Once more the
network splits into two parts the active one, which contributes to the couplings, being formed now by the first $K$ chiral channels. 
An explicit method for computing the  $\xi_k$  is presented in App.~\ref{sec:app_gen_eq_31}. Here for the sake of simplicity we discuss
only the case of  $K=2$. For this choice the matrix $T$ defined in Eq.~\eqref{eq:T_matrix_def} is $2\times 2$ and equal to 
\begin{eqnarray}
T=\left(\begin{array}{cc}
-i\sqrt{1-\tau_1}e^{-i\phi_1} & \sqrt{\tau_1}e^{-i\phi_1}\\
-i\sqrt{\tau_1}e^{-i\phi_2} & -\sqrt{1-\tau_1}e^{-i\phi_2}
\end{array}\right)\;.
\end{eqnarray}
It can be cast in diagonal form $T=UDU^\dag$ with 
\begin{eqnarray}
D=\left(\begin{array}{cc}
e^{i \theta_+}
& 0\\
0 & e^{i \theta_-}
\end{array}\right), \qquad 
U=\left(\begin{array}{cc}
\frac{u_+}{\sqrt{\tau_1+|u_+|^2}}  & \frac{u_-}{\sqrt{\tau_1+|u_-|^2}}\\
\frac{\sqrt{\tau_1} }{\sqrt{\tau_1+|u_+|^2}} & \frac{\sqrt{\tau_1} }{\sqrt{\tau_1+|u_-|^2}}
\end{array}\right) \nonumber , 
\end{eqnarray}
where
\begin{eqnarray}
e^{ i \theta_\pm } &=&\frac{1}{2}\Bigg[-\sqrt{1-\tau_1}\left(ie^{-i\phi_1}+e^{-i\phi_2}\right)\\
\nonumber
&\pm&\sqrt{\left(e^{-i\phi_2}-ie^{-i\phi_1}\right)^2-\tau_1\left(ie^{-i\phi_1}+e^{-i\phi_2}\right)^2}\Bigg], \\
u_{\pm}&=&\sqrt{{1-\tau_1}}\left(e^{-i(\phi_1-\phi_2)}+i\right)\\
\nonumber
&\pm&i\sqrt{{(1-\tau_1)(1-e^{-i2(\phi_1-\phi_2)})-2ie^{-i(\phi_1-\phi_2)}(1+\tau_1)}}\;.
\end{eqnarray}
 From Eq.~\eqref{eq:xi_k_closed} it then follows 
\begin{eqnarray}
\label{eq:k=2}
\xi_k=\frac{u_+e^{ i k \theta_+} -u_-e^{ i k \theta_-}}{u_+-u_-}\;,
\end{eqnarray}
which exhibits an oscillatory behavior in $k$ and a functional dependence  upon the network parameters  which is rather complex 
  (see e.g.  Fig.~\ref{figurenuovissima} where we report the plot of  $|\xi_1|, |\xi_2|, |\xi_3|$ and $|\xi_4|$  in terms of $\tau_1$ and $\phi_2$ for $\phi_1=0$). 
For generic choice of the  settings, while the interactions are still long range, the couplings are no longer uniform  and exhibit a reach variety of behaviours. 
In particular for  $\tau_1=0$, we have $u_-=0$ and $e^{ i \theta_+}= -i e^{-i \phi_1}$  so that Eq.~(\ref{eq:k=2}) exactly reduces to Eq.~(\ref{asdfas}). For the case  $\tau_1=1$ instead we have $e^{ i \theta_\pm}=\pm e^{-i\left(\frac{2\phi_1+2\phi_2+\pi}{4}\right)}$, $u_-=-u_+$, and
\begin{eqnarray}
\label{eq:odd_even_interactions}
\xi_k&=&e^{-ik\left(\frac{2\phi_1+2\phi_2+\pi}{4}\right)} \times  \left\{\begin{array}{ccl}
0 & &\mbox{for $k$ odd}\\
& & \\
1 & &\mbox{for $k$ even,}
\end{array}\right.
\end{eqnarray}
implying that for these settings the odd (even) sites interact only with odd (even) sites. 
By a close inspection one may also notice that Eq.~(\ref{eq:k=2}) simplifies when setting $\phi_2= \phi_1+ \pi/2$ yielding 
\begin{eqnarray}
\label{eq:odd_even_interactions2}
\xi_k&=&e^{-ik\left(\phi_1+\pi/2\right)} \times  \left\{\begin{array}{ccl}
\sqrt{1-\tau_1} & &\mbox{for $k$ odd}\\
& & \\
1 & &\mbox{for $k$ even.}
\end{array}\right.
\end{eqnarray}
Under these conditions the model results in a modification of the
scheme presented in Eq.~(\ref{asdfas}) where now 
we can identify two different species of sites (the odd and the even ones). All the elements of the same species interact uniformly with intensity $1$, while
any two elements belonging to different species interact  with strength  $\sqrt{1-\tau_1}$ (as in the case of~(\ref{asdfas}) the phase $e^{-ik\left(\phi_1+\pi/2\right)}$ is irrelevant
as it can be absorbed by a proper redefinition of the site operators).

Adopting the conditions that led us to (\ref{eq:odd_even_interactions2}) the ME explicitly reads  
\begin{eqnarray} 
\nonumber 
&&\frac{\partial\hat{\rho}}{\partial t}=\sum_{m}\frac{\gamma}{2}\left(2\hat{a}_m\hat{\rho}\hat{a}_m^\dag-\left[\hat{a}_m^\dag\hat{a}_m,\hat{\rho}\right]_{+}\right)\\
&+&\sum_{k\;\text{even}}\sum_m\gamma\left(\hat{a}_m\left[\hat{\rho},a_{m+k}^\dag\right]_{-}+\left[\hat{a}_{m+k},\hat{\rho}\right]_{-}\hat{a}_m^\dag\right)\\
\nonumber
&+&\sum_{k\;\text{odd}}\sum_{m}\gamma\sqrt{1-\tau_1}\left(\hat{a}_m\left[\hat{\rho},a_{m+k}^\dag\right]_{-}+\left[\hat{a}_{m+k},\hat{\rho}\right]_{-}\hat{a}_m^\dag\right),
\end{eqnarray}
while the associated  effective Hamiltonian~(\ref{eq:chiral_hamiltonian}) reads
\begin{eqnarray}
\nonumber
\hat{H}_{eff}=-\frac{i\gamma}{2}\sum_{k\;\text{even}}\sum_m \left(\hat{a}_m \hat{a}_{m+k}^\dag-h.c.\right)\\
-\frac{i\gamma}{2}\sqrt{1-\tau_1}\sum_{k\;\text{odd}}\sum_m \left(\hat{a}_m\hat{a}_{m+k}^\dag-h.c.\right).
\end{eqnarray}
It is also possible to give an analytical expression for the Lindblad operators $\hat{L}_i$'s and for the rates $\gamma_i$'s  entering the standard 
GKSL form representation~(\ref{eq:lindblad}). Following the derivation of Sec.~\ref{STANDARD}  it turns out that, irrespectively from the total number $M$ of sites of the network, there are  only two non-zero terms to consider. 
In particular for $M$ even, setting $i=1,2$   we have 
\begin{eqnarray} \gamma_{i}&=& {M} \; \gamma  \frac{ 1 + (-)^i \sqrt{1-\tau_1}}{2} \;, 
\\
\hat{L}_{i}&=&\frac{1}{\sqrt{M}}\sum_{m=1}^{M}(-1)^{m i } \; \hat{a}_{m}\;. \label{QUESTIQ} 
 \end{eqnarray}  
 For the special case  $\tau_1=1$, due to the degeneracy in the rates  $\gamma_1=\gamma_2 = M\gamma/2$,
we can equivalently replace the operators~(\ref{QUESTIQ}) with 
\begin{eqnarray}
\hat{L}_{1}&=&\sqrt{\frac{2}{M}} \sum_{j=0}^{M/2-1}\hat{a}_{2j+1}\;, \\
\hat{L}_{2}&=&\sqrt{\frac{2}{M}} \sum_{j=1}^{M/2}\hat{a}_{2j}\;,
 \end{eqnarray} 
 which explicitly account for the fact that in this regime the even sites are decoupled from the odd ones. 
Similarly for $M$ odd, we have 
 \begin{eqnarray} \gamma_{i}&=&\gamma M \frac{1 + (-)^i \sqrt{1-(1-1/M^2)\tau_1}}{2}\;, 
\\
\hat{L}_{i}&=&\frac{1}{\sqrt{A}}\Bigg(\sum_{m=0}^{(M-1)/2}\hat{a}_{2m+1}\\
&+&\frac{-1+(-1)^i \sqrt{M^2-(M^2-1)\tau_1}}{(M-1)\sqrt{1-\tau_1}}\sum_{m=1}^{(M-1)/2}\hat{a}_{2m}\Bigg)\;,  \nonumber  \\ 
A&=&\frac{M+1}{2}+\frac{1}{2(M-1)}\left|\frac{1+\sqrt{M^2-(M^2-1)\tau_1}}{\sqrt{1-\tau_1}}\right|^2 \nonumber 
 \end{eqnarray}
which for $\tau_1=1$ reduces to 
 \begin{eqnarray}
 \gamma_{i}&=&\frac{M +(-1)^i}{2}\gamma\;,  \\ 
\hat{L}_{1}&=&\sqrt{\frac{2}{M+1}} \sum_{j=0}^{(M-1)/2}\hat{a}_{2j+1}\;, \\
\hat{L}_{2}&=&\sqrt{\frac{2}{M-1}} \sum_{j=1}^{(M-1)/2}\hat{a}_{2j}\;.
\end{eqnarray}

%%%%%%
%%%%%%%%

\subsection{Retaining only one interaction}
In all the examples discussed so far, the effective couplings exhibit long range interaction terms.  Here we want to show 
 that it is possible to set the transmissivities and the phases of the regular network in order to retain only one kind of interaction among the nodes, for instance only first-neighbor or only second-neighbor interactions.
To this aim, consider the case where we want to nullify all the interactions but the first-neighbor ones, i.e. 
\begin{eqnarray} 
\label{CONDK} 
\xi_k =0 \;, \qquad  \mbox{for all $k\geq 2$.}\end{eqnarray}
As we shall see in what follows, there is an upper bound for the value of the intensity of the coupling constant $\xi_1$ appearing in  Eq.~(\ref{eq:first_neighbor}) for which
the above conditions can be met.
In particular it turns out that this goal cannot 
 be fulfilled for values of $\tau_1$ which are below the  threshold value $3/4$. 
On the contrary for 
 $\tau_1 \in [ 3/4, 1]$, Eq.~(\ref{eq:first_neighbor}) can be enforced with 
 \begin{eqnarray}  \label{FFD}
\xi_1=-ie^{-i\phi_1}\sqrt{1-\tau_1}\;,
\end{eqnarray} 
  by properly tuning the remaining  network phases, and 
 by choosing  the remaining transmissivities according to the recursive formula 
 \begin{eqnarray}
\label{eq:general_trans_amp}
\tau_k=1-\left( \frac{1-\tau_{k-1} \tau_{k-2} \cdots \tau_1}{\tau_{k-1} \tau_{k-2} \cdots \tau_1} \right) \left( \frac{1-\tau_{k-1}}{\tau_{k-1}}\right) \;.
\end{eqnarray}
While the derivation of this result is reported in App.~\ref{sec:app_eq_dem}, a couple of remarks are mandatory: 
\begin{itemize} 
\item As in the case of Eq.~(\ref{QUESTAQUI}) the phase $\phi_1$ appearing in (\ref{FFD}) is irrelevant and can be eliminated by a proper redefinition of the system operators.
\item The solution presented here can be easily adapted to retain only $n^{th}$-neighbour interactions, with $n$ arbitrary integer. In this case it suffices to set all the transmissivities $\tau_k=1$ for $i<n$ and then apply Eq.~\eqref{eq:general_trans_amp} with the index shift $i\rightarrow i+n$.
\end{itemize} 

Under the  conditions~(\ref{FFD}) and (\ref{eq:general_trans_amp}), 
Eq.~(\ref{eq:the_master_equation11}) reduces to 
\begin{eqnarray}
\nonumber
\frac{\partial\hat{\rho}}{\partial t}&=&\sum_{m}\frac{\gamma}{2}\left(2\hat{a}_m\hat{\rho}\hat{a}_m^\dag-\left[\hat{a}_m^\dag\hat{a}_m, \hat{\rho}\right]_{+}\right)\\
&+&\sum_{m}\gamma\sqrt{1-\tau_1}\left(\hat{a}_m\left[\hat{\rho},\hat{a}_{m+1}^+\right]_{-}+\left[\hat{a}_{m+1},\hat{\rho}\right]_{-}\hat{a}_m^\dag\right)\;,  
\nonumber 
\end{eqnarray}
where for the sake of simplicity we set $\phi_1=- \pi/2$. 
This can be casted in standard GKSL form (\ref{eq:lindblad}) with an effective  Hamiltonian~(\ref{EFFECTIVE}) 
\begin{eqnarray}
\hat{H}_{eff}=\hat{H}_1\label{eq:first_neighbor}
=-\frac{i\gamma \sqrt{1-\tau_1}}{2} \; \sum_m\left(\hat{a}_m\hat{a}_{m+1}^\dag-h.c.\right)
\end{eqnarray}
that  contains only first neighbour exchange interactions and, 
as in the model considered in Ref.~\cite{1367-2630-14-6-063014},  exhibits an explicit chiral symmetry which induces a global sign flip when reversing the ordering of the sites 
 $\{S_1,S_2,\cdots S_M\} \rightarrow \{S_M,S_{N-1},\cdots S_1\}$.  
Once more, the associated Lindblad operators $\hat{L}_i$ can  be explicitly computed following the procedure detailed in Sec.~\ref{STANDARD}: 
 in this case however, at variance with the schemes analyzed in Sec.~\ref{sec:examples}, a closed analytical expression for them is less informative  since there are  $M$ collective jump operators each with different weights for the various nodes (this was somehow to be expected, since here the signal is transmitted through as many channels as the nodes).

\section{\label{sec:conclusions}Conclusions}
We have examined a cascade network thanks to which it is possible to simulate a reach variety of dissipative regimes.
In particular by properly setting the system parameters one can achieve configurations where, for an arbitrary number of sites,  only some of them interact, in opposition with the typical case of a linear cascade system where the first node of the cascade interacts with all the subsequent nodes.
We think that this work might open new perspectives on both cascade systems physics and many-body dissipative systems. Indeed, quantum cascade systems have mainly been studied in the context of simple linear chains, and so for the entanglement content of the associated steady states. On the other hand most numerical simulations of many-body open quantum systems are limited to a few nodes because of the complexity of the computation. Thus, experimentally implementing a cascade network as the one described in this paper could represent a new approach for the quantum simulation of such complex systems.
Finally it might be interesting to check whether it is possible to reproduce other dynamical models by changing some parameters of the system, like the interaction between the nodes and environmental modes ({\it e.g.} cubic or quartic interaction Hamiltonians instead of quadratic), the quantum state of the environment ({\it e.g.}\ a squeezed reservoir) or the spatial configuration of the various elements of the network.
\\

We acknowledge the FQXi foundation for financial support in the "Physics of what happens" program.

%%%%%% BIBLIOGRAPHY %%%%%%%%%%%%%

%\looseness=-1
\appendix

\section{\label{STAND} Computing the Lindblad operators of the standard GKSL form} 

A direct application of Ref.~\cite{cusumano_coll_model} provides a way to determine the Lindblad operators $\hat{L}_i$ of (\ref{eq:lindblad}) and their corresponding rates $\gamma_i$.
For this purpose one needs to construct a $2M \times 2M$ Hermitian block matrix 
\begin{eqnarray}
\Omega =\begin{pmatrix}
\Xi_{1,1} &\Xi_{1,2}& \cdots &\Xi_{1,M} \\
\Xi_{2,1}& \Xi_{2,2} & \cdots &\Xi_{2,M} \\
\cdots & \cdots & \cdots & \cdots \\ 
\Xi_{M,1}& \Xi_{M,2} & \cdots &\Xi_{M,M} \\
\end{pmatrix} 
\end{eqnarray} 
 formed by  $2\times 2$ blocks $\{ \Xi_{m,m'}\}_{m,m'=1,\cdots, M}$. Specifically for the master equation~(\ref{eq:the_master_equation}) the diagonal blocks are 
all identical and equal to 
\begin{eqnarray}
\Xi_{m,m} &=&\gamma \begin{pmatrix}
0&0\\
0& 1
\end{pmatrix}\;,
\end{eqnarray} 
while the off-diagonal ones are given by 
\begin{eqnarray} 
\Xi_{m,m'} &=&\Xi_{m',m}^\dag=  \gamma \begin{pmatrix}
0&0\\
0& \zeta_{m,m'} 
\end{pmatrix} \;, 
\end{eqnarray}
for all $m'>m$, with $\zeta_{m'm}$ the coefficients entering Eq.~(\ref{eq:int_term_path}). 
The eigenvalues of $\Omega$ provides now the rates $\gamma_i$. The corresponding $\hat{L}_i$ instead are obtained as the components of 
the column vector 
\begin{eqnarray} \label{DEFL}
\mathbf{L} = W^\dag  \begin{pmatrix}
\hat{a}_1^\dag \\
\hat{a}_1 \\ 
\hat{a}_2^\dag \\
\hat{a}_2 \\ 
\vdots 
\end{pmatrix} \end{eqnarray} 
where $W$ is the $2M \times 2M$ unitary operator which diagonalizes $\Omega$, i.e. 
$WD W^\dag = \Omega$ with $D= \mbox{diag}[\gamma_i]$. The whole construction can be further simplified as reported in the Eqs.~(\ref{THETA1})-(\ref{DEFLf}) 
by  noticing that the odd rows of $\Omega$ (as well as its odd columns) contains only zero elements and can be hence neglected (their associated eigenvalues being null). Removing them transform $\Omega$ into the $M\times M$ matrix $\Theta$
and (\ref{DEFL}) into (\ref{DEFLf}). This is 
a direct consequence of the fact that, at variance with the cases addressed in Ref.~\cite{cusumano_coll_model}
 in model we are studying here, all the environmental modes entering the network are intialized at zero temperature.

\section{\label{sec:app_gen_eq_31}Closed formula for the $\xi_k$}

%%%%%%%%%%
\begin{figure}[!t]
\centering
\includegraphics[scale=0.8]{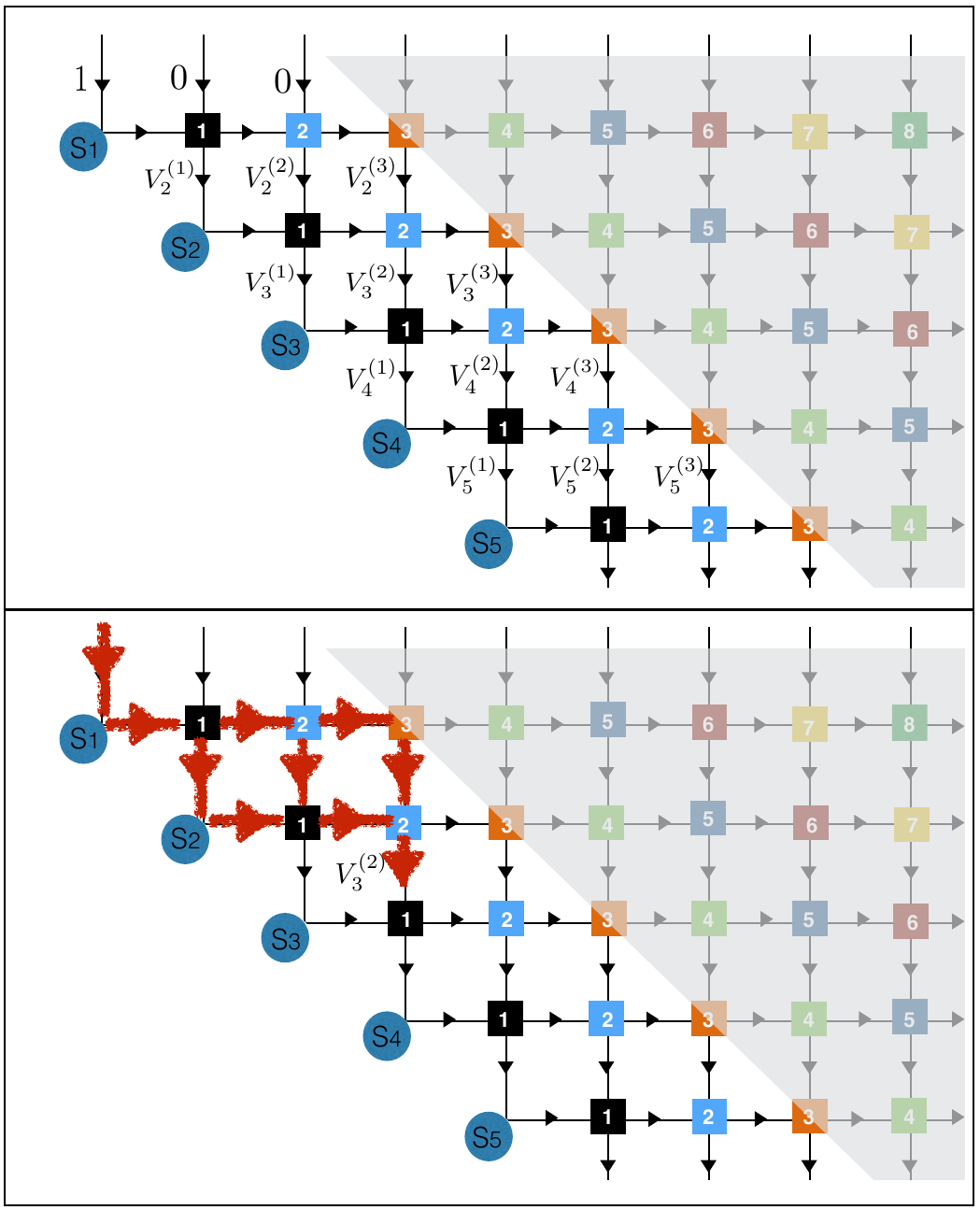}
\caption{Pictorial representation of the components of the vectors $\vec{V}_\ell$ of Eq.~(\ref{DEFA}) for the cases $K=3$: for $\ell$ and $k$ integer,
 $V_{\ell}^{(k)}$ describes the amplitude probability associated with the path connecting the first entry of level $1$ to the $k$-th entry of the level $\ell$. In the lower panel we enlightened the paths
that contribute to the amplitude probability that yields the value of $V_3^{(2)}$.}
\label{fig:A_vector_scheme}
\end{figure}
%%%%%%%

In this section we present the explicit derivation of the coupling strengths~(\ref{eq:xi_k_def}) for a regular network with transmissivities $BS_{m,m+K}=0$ for given value of $K\geq 2$. 
For this purpose let us introduce the $K$ dimensional, column  vector $\vec{V}_\ell$ whose components represent the amplitude probabilities 
associated with the propagation of the mode ${\cal E}^{(1)}$ from level $1$ to the first $K$ entries of level $\ell$, i.e.
\begin{eqnarray} \label{DEFA} 
(\vec{V}_\ell)^T &:=& (V_{\ell}^{(1)}, V_{\ell}^{(2)},\cdots,  V_{\ell}^{(K)})\;,  \quad \mbox{$\ell \geq 2$}\;,
\\
(\vec{V}_1)^T&:=&(1, 0,\cdots,  0)\;,
\end{eqnarray} 
see Fig.~\ref{fig:A_vector_scheme}.
It is important to observe that in this notation the  amplitude $A_{k+1 \leftarrow 1}^{(1)}$, through which, according to Eq.~(\ref{eq:xi_k_def}), the coupling constant $\xi_k$ is expressed, corresponds to the first entry of the vector $\vec{V}_{k+1}$, i.e. 
\begin{eqnarray}  \label{NOTATION1} 
A_{k +1\leftarrow 1}^{(1)} = V_{k+1}^{(1)} = (\vec{V}_1)^T \cdot\vec{V}_{k+1}\;. 
\end{eqnarray} 
One notices also 
 that  $\vec{V}_{\ell +1}$ and $\vec{V}_\ell$ are related as \begin{eqnarray}
\label{eq:recursion_A_vector}
\vec{V}_{\ell +1}=T \vec{V}_\ell \;, 
\end{eqnarray}
where $T$ is the $K\times K$ unitary matrix which rules the propagation of signals from one level of the network to the next, its elements being 
\begin{widetext}
\begin{eqnarray}
\label{eq:T_matrix_def}
T_{ij}=\left\{\begin{array}{ccl}
e^{-i\phi_i}\sqrt{\tau_i} & &\mbox{for}\;j=i+1\\
& & \\
e^{-i\phi_i}(-i\sqrt{1-\tau_i})(-i\sqrt{1-\tau_{i-1}})& &\mbox{for}\;j=i\\
& & \\
e^{-i\phi_i}(-i\sqrt{1-\tau_{j-1}})\left(\displaystyle\prod_{\ell=j}^{i-1}\sqrt{\tau_{\ell}}\right)(-i\sqrt{1-\tau_i})& &\mbox{for}\;j<i\\
& & \\
0 & & \mbox{otherwise}
\end{array}\right.
\end{eqnarray}
\end{widetext}
where, by convention, we make the substitution $-i\sqrt{1-\tau_0}\rightarrow1$ wherever necessary.
By expressing $T$ in diagonal form and by iterating  Eq.~\eqref{eq:recursion_A_vector} we can then write:
\begin{eqnarray}
\vec{V}_{\ell+1} =T^\ell \vec{V}_1
=UD^\ell U^{\dag}\vec{V}_1 \;, 
\end{eqnarray}
where $D$ is a diagonal matrix formed by the eigenvalues of $T$ and $U$ is the unitary matrix formed by the corresponding eigenvectors, i.e.  $T = UDU^\dag$.
Therefore, remembering (\ref{eq:xi_k_def}) and (\ref{NOTATION1}) we have 
\begin{eqnarray}
\xi_k= A_{k+1\leftarrow 1 }^{(1)} = 
(\vec{V}_1)^T \cdot\vec{V}_{k+1} 
&=&(\vec{V}_1)^T  \cdot\left[U D^k U^{\dag}\right] \vec{V}_1 \;. \nonumber \\ 
\label{eq:xi_k_closed}
\end{eqnarray}

\section{\label{sec:app_eq_dem}Proof of Eq.~\eqref{eq:general_trans_amp}}
\subsubsection{Derivation of Eq.~\eqref{eq:general_trans_amp}} 
\begin{figure}[!t]
\centering
\includegraphics[scale=0.60]{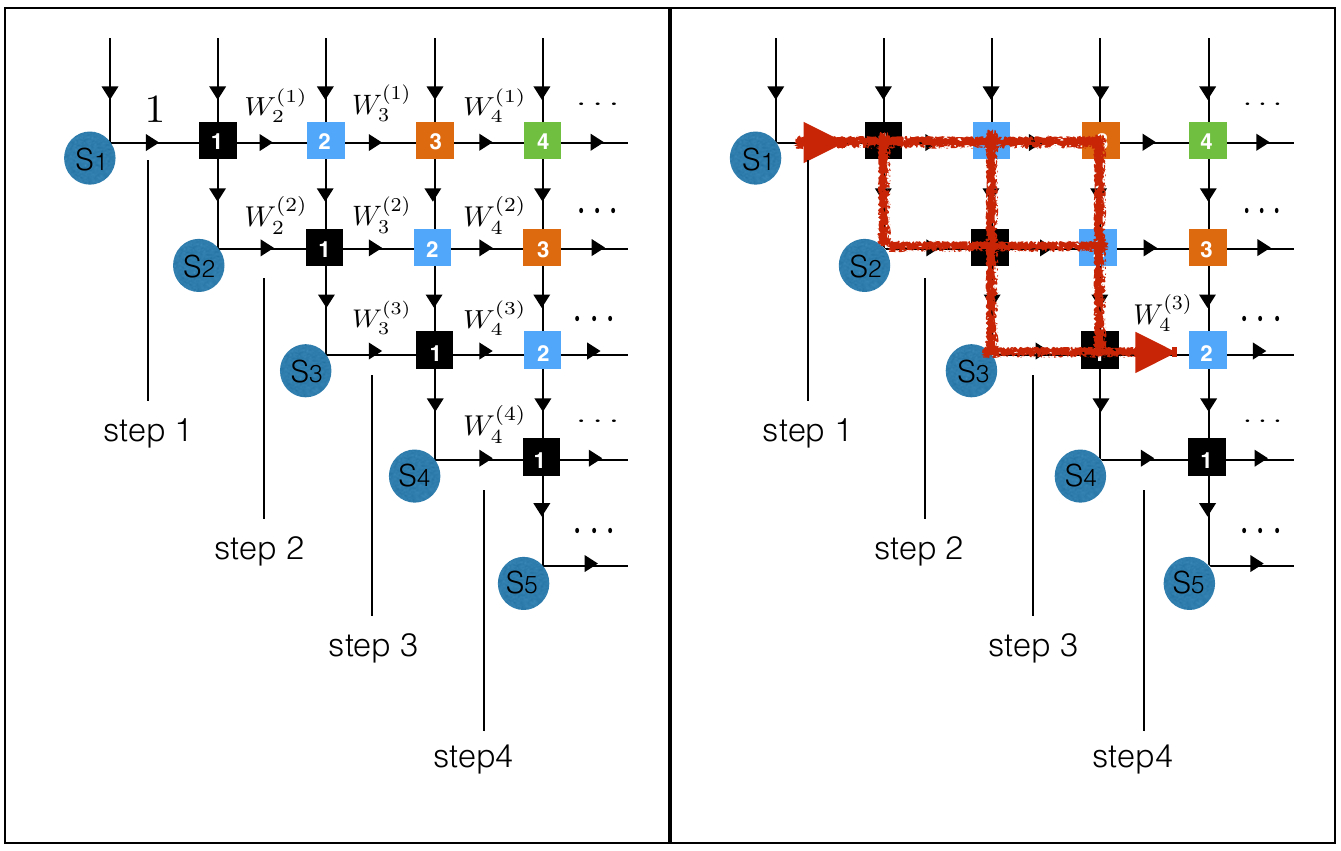}
\caption{Illustration of the amplitudes $W_\ell^{(k)}$ which label the horizontal lines  of the network: for $\ell$ and $k$ integer they represent the probability amplitude
associated with the propagation of a signal  emerging from  $S_1$ and reach the $\ell$-th level  of the $k$-th horizontal step of the network. As an example in the left panel we
enlightened the paths which enter in the definition of $W_{4}^{(3)}$.} 
\label{fig:gamma_coeff}
\end{figure}

In order to prove the formula~(\ref{eq:general_trans_amp}) we find it useful to adopt the notation  presented in Fig.~\ref{fig:gamma_coeff}. Here, at variance with what we have done  in Sec.~\ref{sec:examples},
we now  label the horizontal elements of the network. In particular, for $\ell$ and $k$ integers,  we use the symbol $W_{k}^{(\ell)}$ to indicate the amplitude of the
signal reaching the $k$-th horizontal step  of the $\ell$-th level starting from  $S_1$. They are  connected through the action of the network beam splitters via a series of linear equations which we report here for the first values of $k$, i.e. 
\begin{eqnarray} 
W_2^{(1)} &=& \sqrt{\tau_1}\;, \qquad W_{2}^{(2)} = - i \sqrt{1 -\tau_1}e^{-i\phi_1}\;; \label{PRIMA} \\
W_3^{(1)} &=& \sqrt{\tau_2} \; W_2^{(1)} \;, \label{SECONDA} \\
 W_{3}^{(2)} &=& (- i \sqrt{1 -\tau_1}e^{-i\phi_1}) (- i \sqrt{1 -\tau_2}e^{-i\phi_2}) W_2^{(1)} 
\nonumber \\
&&+ \sqrt{\tau_1} \; W_2^{(2)} \;, \label{TERZA} \\
W_{3}^{(3)} &=& (\sqrt{\tau_1}e^{-i\phi_1}) (-i \sqrt{1-\tau_2} e^{-i\phi_2}) W_2^{(1)} 
\nonumber \\
&&+ (-i \sqrt{1-\tau_1} e^{-i\phi_1}) W_2^{(2)} \;.
\end{eqnarray}  
In this notation the probability amplitude  $A_{k\leftarrow 1}^{(1)}$
corresponds to  the element $W_{k}^{(k)}$, and
therefore, thanks to Eq.~(\ref{eq:xi_k_def}) we can express the coupling constants as 
\begin{eqnarray}  \label{NOTATION2} 
\xi_k = A_{k +1\leftarrow 1}^{(1)} = W_{k+1}^{(k+1)}\;. 
\end{eqnarray} 
We also point out that for $\ell=1$ and $k$ generic, the following identity holds 
\begin{eqnarray} 
W_k^{(1)} &=& \sqrt{\tau_{k-1} \tau_{k-2} \cdots \tau_2 \tau_1}\;,
 \label{SECONDAAA}
\end{eqnarray} 
which allows us to rewrite (\ref{eq:general_trans_amp}) as 
 \begin{eqnarray}
\label{eq:general_trans_ampNEW}
\tau_k=1-\left( \frac{1-|W_k^{(1)}|^2}{|W_k^{(1)}|^2} \right) \left( \frac{1-\tau_{k-1}}{\tau_{k-1}}\right) \;.
\end{eqnarray}

%%%%%%%%
%%%%%%

Consider then the condition~(\ref{CONDK}) for $k=2$ which ensures the nullification of the 
second-neighbour interaction constant. According to (\ref{NOTATION2}) imposing $\xi_2=0$ means to set the system parameters in such a way that $W_{3}^{(3)}$ nullifies,  i.e. 
\begin{eqnarray} 
\label{PPPP} 
 \sqrt{\tau_1}  (-i \sqrt{1-\tau_2}  e^{-i \phi_2} ) 
W_2^{(1)}  
 + (- i \sqrt{1-\tau_1} ) W_2^{(2)}   =0 \nonumber \;,
\end{eqnarray} 
that implies  
\begin{eqnarray} 
\phi_2&=&\phi_1+\pi/2\;, \label{QUS1} \end{eqnarray} 
and 
\begin{eqnarray} 
 \tau_2 &=&  1- \left( \frac{1- |W_2^{(1)}|^2}{|W_2^{(1)}|^2} \right) \left( \frac{1 -\tau_1}{\tau_1} \right) \;,  \label{QUS} 
\end{eqnarray} 
where we used the fact that the amplitudes $W_2^{(1)}$ and $W_2^{(2)}$ are complementary (i.e. their square modulus sum up to 1) and this 
proves the validity of the formula~(\ref{eq:general_trans_ampNEW}) for $k=2$. 

\begin{figure}[!t]
\centering
\includegraphics[scale=0.75]{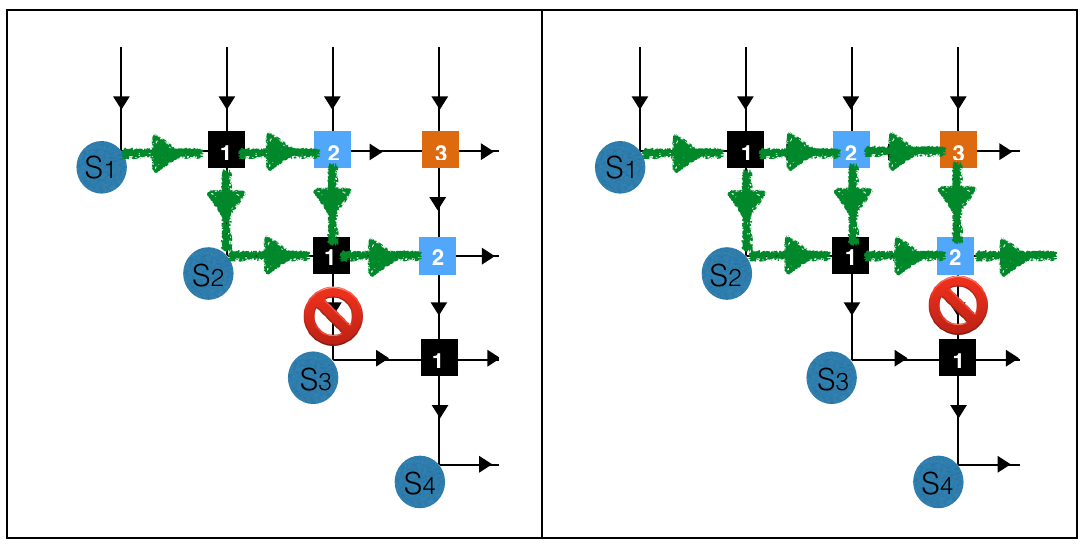}
\caption{Left Panel: scheme of the possible paths when eliminating second-neighbour interactions, i.e. nullifying $\xi_2$. The signal coming from $S_1$ (green path) splits away in the first beam splitter, following two distinct paths. They then recombine in the beam splitter with $\tau_1$at level 2 of the network, just above $S_3$, both ending up in the channel propagating on the right thanks to interference effects. Right panel: the same as in the left panel, but for third-neighbour interactions. Here we notice that, once we have eliminated second-neighbour interactions, no signal coming from $S_1$ can make its way up to $S_3$, i.e. no signal arrives at level 3 of the network. This implies that all the signals coming from $S_1$ must end up entirely in channel $\mathcal{E}_2$ in order to eliminate also third-neighbour interactions.}
\label{fig:iteration_scheme}
\end{figure}

Once these conditions are met, the signal from $S_1$ reaches the step 3 of the interferometer without touching $S_3$ but populating only the first two levels of the network
producing there amplitudes $W_3^{(1)}$ and $W_3^{(2)}$. The explicit values of these quantities can be computed as  in Eqs.~(\ref{SECONDA}) and (\ref{TERZA}), yet
for what concern to us it is sufficient to observe that due to the conservation of probability and the condition $W_3^{(3)}=0$,  it follows that also these two amplitudes have to be
 complementary, i.e. 
\begin{eqnarray} \label{imo} 
W_3^{(2)} =   e^{-i \alpha_3} \; \sqrt{1 - |W_3^{(1)}|^2} \;, 
\end{eqnarray} 
with $\alpha_3$ being an irrelevant phase. With this observation in mind  let us now consider  the condition~(\ref{CONDK}) with $k=3$. 
Again to enforce it we must prevent signals to reach $S_4$ by setting $W_4^{(4)}=0$. 
In this case however we notice that since
 we have already imposed $W_3^{(3)}=0$, the beam splitter of transmissivity $\tau_1$ on the third level
has no horizontal input that can be used to destructively interfere with a possible vertical signal that reaches it. Hence to have null value of $\xi_3$ we must have once more that 
all the signals from $S_1$ remain confined into the first two levels of the network, i.e. 
\begin{eqnarray} 
\label{PPPP2} 
 \sqrt{\tau_2}  (-i \sqrt{1-\tau_3}  e^{-i \phi_3} ) 
W_3^{(1)}  
 + (- i \sqrt{1-\tau_2} ) W_3^{(2)}   =0 \nonumber \;,
\end{eqnarray} 
see right panel of Fig.~\ref{fig:iteration_scheme}.
Exploiting Eq.~(\ref{imo})  this reduces to the following conditions for $\tau_3$ and $\phi_3$
 \begin{eqnarray} 
\phi_3&=& \alpha_3 + \pi\;,  \\ 
 \tau_3 &=& 1- \left( \frac{1- |W_3^{(1)}|^2}{|W_3^{(1)}|^2} \right) \left( \frac{1 -\tau_2}{\tau_2} \right) \;,\end{eqnarray} 
 that represent the $k=3$ counterparts of Eqs.~(\ref{QUS1}) and (\ref{QUS}) respectively, the second
 corresponding also to  (\ref{eq:general_trans_ampNEW})  for $k=3$. 
 The same reasoning can now be iterated  to $k=4$:  indeed having imposed $\xi_2=\xi_3=0$ forces the signals to reach the fourth step of the network
 by only populating the first two levels with complementary intensities $W_4^{(1)}$ and $W_4^{(2)}$ which have to fulfil the condition
 \begin{eqnarray} 
\label{PPPP2e} 
 \sqrt{\tau_3}  (-i \sqrt{1-\tau_4}  e^{-i \phi_4} ) 
W_4^{(1)}  
 + (- i \sqrt{1-\tau_3} ) W_4^{(2)}   =0 \nonumber \;,
\end{eqnarray} 
that is 
 \begin{eqnarray} 
\phi_4&=& \alpha_4 + \pi\;,  \\ 
 \tau_4 &=& 1- \left( \frac{1- |W_4^{(1)}|^2}{|W_4^{(1)}|^2} \right) \left( \frac{1 -\tau_3}{\tau_3} \right) \;,\end{eqnarray} 
 and so on. 
 %%%%%%%%%%%%%%%%%%%%%%%%%%%%%%%%%
%%%%%%%%%%%%%%%%%%%%%%%%
 \subsubsection{\label{sec:demonstration} Range of applicability of Eq.~\eqref{eq:general_trans_amp}}
By recursion on the various level, Eq.~\eqref{eq:general_trans_amp} induces a functional dependence of all the transmissivities of the network upon $\tau_1$. For instance for $k=2,3,4$  we get 
\begin{eqnarray} 
&\tau_2(\tau_1) :=  1 - \frac{(1-\tau_1)^2}{\tau_1^2}\;, \quad \tau_3(\tau_1) = 1 - \frac{(1-\tau_1)^3}{(2\tau_1-1)^2}\;,& \nonumber \\
&\tau_4(\tau_1) = 1 - \frac{(1-\tau_1)^4}{(\tau_1^2+\tau_1-1)^2}\;.& \label{Esempi} 
\end{eqnarray} 
These expressions produce legitimate values of the transmissivities only for a limited range of values of $\tau_1$:
for instance for $k=2$ one has that $\tau_2(\tau_1)  \in [0,1]$ iff the transmissivity $\tau_1$ is larger than $0.5$; for  $k=3$ instead $\tau_3(\tau_1)  \in [0,1]$ 
iff $\tau_1$ is larger than $(\sqrt{5}-1)/2\simeq 0.618$; while finally for $k=4$ instead $\tau_4(\tau_1)  \in [0,1]$ 
iff $\tau_1$ is larger than ${2}/{3}$.

%\begin{center}
%\begin{tabular}{|l|}
%\hline
%$\tau_2 (\tau_1)  \in [0,1]  \Longrightarrow \tau_1 \geq 1/2=0.5$  \\
%$\tau_3(\tau_1)  \in [0,1]  \Longrightarrow  \tau_1 \geq (\sqrt{5}-1)/2\simeq 0.62$  \\ 
%$\tau_4(\tau_1)  \in [0,1]  \Longrightarrow  \tau_1 \geq {2}/{3}\simeq 0.67$ \\
%$\tau_6(\tau_1)  \in [0,1]  \Longrightarrow\tau_1\geq {1}/{\sqrt{2}}\simeq 0.71$ \\ 
%$\tau_{10}(\tau_1)   \in [0,1]  \Longrightarrow\tau_1\gtrsim0.73$ \\ 
%$\tau_{15}(\tau_1)   \in [0,1]  \Longrightarrow\tau_1\gtrsim0.74$ \\ 
%\hline
%\end{tabular}
%\end{center} 
A better insight on the problem can be reached by noticing the following fact 
 associated with formula~\eqref{eq:general_trans_amp}:
\begin{enumerate}  
\item
for all $\tau_1\in [0,1]$, all the functions $\tau_k(\tau_1)$ are upper bounded by $1$. For $k=2,3,4$ this can be easily established by looking at~(\ref{Esempi}).
For arbitrary $k$ instead the thesis follows by exploiting the fact that for all $k$, $1- \tau_{k+1}(\tau_1)$ and $1- \tau_{k-1}(\tau_1)$ must always have the same sign as it can be
easily verified by looking at 
 the identity 
\begin{eqnarray}
1- \tau_{k+1} = \left(\frac{1 - \tau_{k-1} \tau_{k-2} \cdots \tau_1}{1-  \tau_{k-1} \tau_{k-2} \cdots \tau_1 -\tau_{k-1}}\right)^2 \; (1 -\tau_{k-1})\;, \nonumber \\\label{TAU134} 
\end{eqnarray}
derived via a simple iteration of 
Eq.~(\ref{eq:general_trans_amp}).
Also one may notice that  for $\tau_1=1$ we have $\tau_k(\tau_1=1) = 1$ for all $k$;

\item let   $\bar{\tau}_1\in [0,1]$ such that   $\tau_k(\bar{\tau}_1) \geq 0$ for all $k$. Then the same inequalities holds for all transmissivities  $\bar{\bar{\tau}}_1$ which are larger than $\bar{\tau}_1$. This fact can be established by observing that the r.h.s. of Eq.~(\ref{eq:general_trans_amp}) that defines $\tau_k$ in terms of the transmissivities of lower order, is an increasing function of all the parameters $\tau_{k-1}$, $\tau_{k-2}$, $\cdots$, $\tau_1$. Specifically, for $k=2$ this implies that $\tau_2(\bar{\bar{\tau}}_1) \geq \tau_2(\bar{\tau}_1)\geq 0$. For $k=3$ instead we have $\tau_{3}(\bar{\bar{\tau}}_1,\bar{\bar{\tau}}_2) \geq \tau_{3}(\bar{\tau}_1,\bar{\tau}_2)\geq 0$ where $\bar{\bar{\tau}}_2= \tau_2(\bar{\bar{\tau}}_1)$;
\item for $\tau_1=3/4$, Eq.~(\ref{eq:general_trans_amp}) yields  
\begin{eqnarray}
\label{eq:general_trans_amp_3/4}
\tau_k(\tau_1=3/4)=1-\frac{1}{(k+1)^2}=\frac{k(k+2)}{(k+1)^2},
\end{eqnarray}
which are legitimate transmissivities for all $k$. 
A prove of this fact can be easily obtained by induction. 
\end{enumerate} 
Putting together these observations  we can then arrive to the conclusion that 
\begin{eqnarray} 
\tau_1\geq 3/4 \; \Longrightarrow \tau_k(\tau_1)  \in [0,1]  \;\; \mbox{for all $k$.} 
\end{eqnarray}  
Numerical evidence suggests that $\tau_1\geq 3/4$ is also a necessary condition for the applicability of the formula~(\ref{eq:general_trans_amp}) (for instance by explicitly 
evaluating $\tau_{k}(\tau_1)$ for $k=10$ we have that $\tau_1$ cannot be smaller than $0.74$).


\begin{thebibliography}{99}
\bibitem{NOTA}
It is worth commenting on the fact that Eq.~(\ref{DEF11}) makes it explicit the connection between the collisional approach to QCS 
of Ref.~\cite{prl108040401,jpbatmolopt45154003,cusumano_coll_model} and the more traditional derivation of Refs.~\cite{PhysRevLett.70.2273,PhysRevA.50.1792,PhysRevA.31.3761,PhysRevLett.70.2269,gardinerbook}. Indeed the field operators $\hat{c}_{\mathcal{E}^{(m')}_n}$ and $\hat{c}_{\mathcal{E}^{(m)}_n}$ appearing on the right-hand-side
of such equation, are nothing but the input noise operators acting on the subsystems $S_m'$ and $S_m$ respectively.
\end{thebibliography}
\end{document}